# Theoretical model of viscous friction
# inside steadily sheared foams and concentrated emulsions

S. Tcholakova,[1] N. D. Denkov,[1,*] K. Golemanov,[1] K.P. Ananthapadmanabhan,[2] A. Lips[2]

[1]*Laboratory of Chemical Physics & Engineering, Faculty of Chemistry, Sofia University,*
*1 James Bourchier Ave., 1164 Sofia, Bulgaria*
[2]*Unilever Global Research Center, Trumbull, Connecticut 06611, USA*

**Abstract**

In a recent letter (Denkov et al., *Phys. Rev. Lett.*, 2008, *in press*) we calculated theoretically the macroscopic viscous stress of steadily sheared foam/emulsion from the energy dissipated inside the transient planar films, formed between neighboring bubbles/drops in the shear flow. The model predicts that the viscous stress in these systems should be a proportional to $Ca^{1/2}$, where Ca is the capillary number and $n = 1/2$ is the power-law index. In the current paper we explain in detail our model and develop it further in several aspects: First, we extend the model to account for the effects of viscous friction in the curved meniscus regions, surrounding the planar films, on the dynamics of film formation and on the total viscous stress. Second, we consider the effects of surface forces (electrostatic, van der Waals, etc.) acting between the surfaces of the neighboring bubbles/drops and show that these forces could be important in emulsions, due to the relatively small thickness of emulsion films (often comparable to the range of action of the surface forces). In contrast, the surface forces are usually negligible in steadily sheared foams, because the dynamic foam films are thicker than the extent of surface forces, except for foams containing micrometer-sized bubbles and/or at very low shear rates. Third, additional consideration is made for bubbles/drops exhibiting high surface viscosity, for which we demonstrate an additional contribution to the macroscopic viscous stress, created by the surface dissipation of energy. The new upgraded model predicts that the energy dissipation at the bubble/drop surface leads to power-law index $n < 1/2$, whereas the contribution of the surface forces leads to $n > 1/2$, which explains the rich variety of foam/emulsion behaviors observed upon steady shear. Various comparisons are made between model predictions and experimental results for both foams and emulsions, and a very good agreement is found.

# 1. Introduction.

Foams and concentrated emulsions, in which the volume fraction of the dispersed phase, $\Phi$, is higher than the volume fraction of closely packed spheres, $\Phi_{CP}$, exhibit complex elasto-visco-plastic rheological behavior, involving non-trivial jamming/unjamming transitions and bubble/drop deformations, which are controlled by the interplay of viscous drag and capillary pressure [1-20]. Partially, this complex behavior is related to the fact that relatively thin liquid films are formed between the dispersed entities (bubbles or drops), due to their confinement in the dispersion. During foam and emulsion flow, high local shear rates are created in these films (compared to the macroscopic shear rate of the dispersion), which results in relatively high viscous dissipation of energy. The bubble/drop deformability leads to non-linear dependences of the film thickness and of the resulting flow characteristics of foams and emulsions on the applied shear stress, and these dependences are still poorly understood.

The rheological properties of foams and concentrated emulsions, subject to steady shear deformation, are usually described by the Herschel-Bulkley model, which includes three parameters – yield stress $\tau_0$, power-law index $n$, and consistency, $k$ [1,4-6,9,11]:

$$\tau = \tau_0 + \tau_V(\dot{\gamma}) = \tau_0 + k\dot{\gamma}^n \qquad (1)$$

Here $\dot{\gamma}$ is the applied shear rate, $\tau$ is total shear stress, and $\tau_V(\dot{\gamma})$ is rate-depending part of this stress. The dependence of the yield stress, $\tau_0$, on foam/emulsion characteristics, such as bubble/drop size, volume fraction of the dispersed phase, and interfacial tension were studied in detail both theoretically and experimentally [1-4,6,9-11] and will not be considered here.

Recently, several research groups demonstrated that, close to the yielding transition, the shear rate inside sheared foams and emulsions could be non-homogenous with co-existence of jammed and unjammed zones [11,13,20,22-25]. Even in these cases, it was found that the rate-dependent fraction of the shear stress could be described by a power-law function, similar to the second term in the right-hand-side of Eq. (1).

The focus of the present study is on this term, $\tau_V(\dot{\gamma})$, which is much less understood in comparison with $\tau_0$. In many studies, the experimental results were described by empirical power-law fits with $n \approx 1/2$ [1,4,18,25-29]. However, in other studies different power-law indexes were reported, without clear understanding how the values of $n$ and $k$ depend on the specific system parameters [9,26,30-31]. Several theoretical models were proposed in literature to describe the viscous friction in steadily sheared foams and concentrated emulsions [1,5,8,32-35], however, neither of these models succeeded to encompass the results, obtained in the various experimental studies.

In addition, recent rheological experiments with foams [26] and observations of the dynamics of single Plateau borders [36] revealed a very strong effect of the type of surfactant used. For example, power-law indexes $n \approx 0.40$ and $n \approx 0.25$ were reported for foams, which were stabilized by surfactants exhibiting low and high surface modulus, respectively, with the viscous stress being much higher for the second type of surfactants [26]. These results demonstrated that the surface dissipation of energy could be significant for systems characterized by high surface viscosity of the bubbles/drops, but no theoretical model of this effect has been proposed so far.

In a recent letter [37], we developed a theoretical model of the viscous friction in steadily sheared, concentrated foams and emulsions with $\Phi > \Phi_{CP}$. The basic assumption in this model was that the viscous dissipation occurs predominantly in the thin films, formed between the neighboring drops and bubbles, which pass by each other dragged by the flow, see Fig. 1. The processes of film formation and thinning were explicitly considered and used to calculate the energy dissipation inside the films. This model predicts power-law index $n \approx 1/2$ and allows one to calculate the consistency, $k$, which was in a very good agreement with experimental data obtained by different groups [4,26,38]. However, the model could not explain the experimental results, in which power-law indexes, $n \neq 1/2$, are determined.

The main purpose of the current paper is to extend further the model from Ref. [37], taking into account several additional effects, which might explain the reach variety of foam/emulsion behaviors observed upon steady shear. As shown below, the upgraded model predicts that the energy dissipation at the bubble/drop surface could explain the results with $n < 1/2$, whereas the surface forces acting in the foam/emulsion films could explain the results with $n > 1/2$. Note that we are interested mostly in the shear rates characterizing the foam/emulsion transportation (ca. 0.5 to 200 $s^{-1}$). Therefore, we assume that the bubble size and volume fraction are known, and do not consider the processes of Ostwald ripening and water drainage, which might be important at longer time-scales [11,39-47].

The paper is structured as follows: In section 2 we describe the various assumptions and consecutive steps, made in the development of the original model from Ref. [37]. In addition, in section 2.3 we upgrade the model by accounting for the possible contribution of the surface forces acting between the neighboring drops/bubbles, and analyze in which cases this contribution is important. In section 3 we consider the possible effects of the viscous friction in the curved meniscus regions, surrounding the films. In section 4 we present the results from numerical calculations. In section 5 we compare the model predictions with experimental results at negligible effect of surface dissipation. The latter effect is considered in section 6 and, finally, we summarize the conclusions in section 7.

**2. Theoretical model.**

For brevity, in the model formulation we discuss explicitly only bubbles in steadily sheared foams. However, the model is equally applicable to concentrated emulsions, and its predictions are compared in section 5 with experimental results for both foams and emulsions.

## 2.1. Main characteristics of deformed bubbles in static foam.

Let us consider first an idealized static foam, consisting of monodisperse bubbles with given volume, $V_B = (4/3)\pi R_0^3$, and volume fraction, $\Phi$, which are arranged in a regular fcc-lattice. The geometrical characteristics of the deformed bubbles in such regular foam could be estimated as follows (see Fig. 1):

The average distance between the geometrical centers of two neighboring bubbles, $l_S$, is proportional to the bubble radius, $R_0$, and could be found from geometrical relations, after assuming a certain model shape of the bubbles. For example, assuming that the bubbles have the shape of rhombic dodecahedrons, one can use the fact that the height of the rhombic pyramid, formed between dodecahedron center and one of the rhombs on the dodecahedron wall, is equal to $l_S/2$. The respective geometrical calculations lead to $l_S \approx 1.812 R_0$ in the limit of a dry foam ($\Phi \to 1$). On the other hand, the decrease of air volume fraction, $\Phi$, at fixed relative arrangement of the bubbles, corresponds to an increase of the center-to-center distance by a factor of $\Phi^{1/3}$, which means that for arbitrary $\Phi > \Phi_{CP}$:

$$l_S \approx \frac{1.812}{\Phi^{1/3}} R_0 \qquad \text{(rhomboic dodecahedron)} \qquad (2)$$

For other model bubble shapes, the functional dependence of $l_S$ on $R_0$ and $\Phi$ is the same, whereas the numerical factor would differ slightly (within a few percent).

For the following consideration, it is convenient to introduce an effective radius of curvature of the deformed bubbles, $R_{eff}^2 = R_{FS}^2 + l_S^2/4$, which is defined as the radius of a spherical surface with just one planar film (instead of the 12 films of the packed rhombic dodecahedra) that would exhibit the same ratio $R_{FS}/l_S$, as the deformed polyhedral bubbles in the static foam, Fig. 2(a). Here $R_{FS}$ is the radius of the foam film between two neighboring bubbles in the static foam. The advantage of using $R_{eff}$, when considering the dynamic collisions between neighboring bubbles in sheared foam, is that one can replace the actual polyhedral bubbles by "imaginary" spherical bubbles with radius of curvature, $R_{eff}$, which have just one planar film in the zone of bubble-bubble contact under consideration. To estimate $R_{eff}$ we have assumed that the film thickness $h \ll l_S$ - this assumption is justified for all systems of interest, and is used consistently throughout the consideration below.

The scaled film radius in the static foam, $R_{FS}/R_0$, could be determined as a function of air volume fraction, by using an expression derived by Princen [1]:

$$S_F(\Phi) = S_1 f(\Phi)/\Phi^{2/3} \qquad (3)$$

where $S_F(\Phi)$ is the area of the bubble surface, occupied by films; $S_1 \approx 1.1053 S_0$ is the area of the deformed bubble at $\Phi \to 1$; $S_0 = 4\pi R_0^2$ is the surface area of non-deformed bubbles; and $f(\Phi)$ is the fraction of wall surface, confining the foam, that is contacted by flattened bubble surface (i.e., the fraction of wall occupied by wetting films). By definition, $S_F = p\pi R_{FS}^2$, where $p$ is the number of films per bubble in the static foam ($p = 12$ for the assumed fcc-structure), which leads to:

$$R_{FS} \approx 0.607 \left[ f(\Phi)/\Phi^{2/3} \right]^{1/2} R_0 \qquad \text{(rhomboic dodecahedron)} \qquad (4)$$

For determining the dependence $f(\Phi)$ we used the relation [1,48-50]:

$$f(\Phi) = P_{OSM}(\Phi)/P_C(\Phi) \qquad (5)$$

where $P_{OSM}$ is the osmotic pressure of the concentrated emulsion/foam and $P_C$ is the capillary pressure of the respective deformed drops/bubbles. On its turn, $P_C$ could be expressed from another rigorous thermodynamic relation [1,48-50]:

$$P_C(\Phi) = \frac{\sigma}{R_0} \left( \frac{\tilde{P}_{OSM}(\Phi)}{\Phi} + 2\frac{S(\Phi)}{S_0} \right) \qquad (6)$$

where $\sigma$ is interfacial tension and $\tilde{P}_{OSM} = P_{OSM}/(\sigma/R_0)$ is dimensionless osmotic pressure. For typical polydisperse emulsion, Princen and Kiss [49] found that the following empirical functions could describe the experimental data for osmotic pressure:

$$\tilde{P}_{OSM}(\Phi) = 0.237\left(\frac{\Phi - 0.715}{1-\Phi}\right) - 0.068\ln\left(\frac{0.285}{1-\Phi}\right) - 0.098 \qquad 0.715 < \Phi < 0.90 \qquad (7a)$$

$$\tilde{P}_{OSM}(\Phi) = \frac{0.00819\Phi^2}{(1-0.9639\Phi)^2} \qquad 0.90 < \Phi < 0.99 \qquad (7b)$$

In Eq. (6), $S(\Phi)$ is the total surface area of the deformed bubble/drop, which can be found by the relation $S/S_0 = 1 + \int_{\Phi_{CR}}^{\Phi} (\tilde{P}_{OSM}/3\Phi^2) d\Phi$. Explicit expressions for $S(\Phi)$ corresponding to Princen's functions, are given by Eqs. (49)-(50) in Ref. [1].

For monodisperse foams of equally sized and regularly arranged bubbles, Hoehler et al. [51] proposed the following empirical relation to describe their experimental data:

$$\tilde{P}_{OSM}(\Phi) = 7.3\frac{(\Phi - 0.74)^2}{(1-\Phi)^{1/2}} \qquad 0.74 < \Phi < 0.99 \qquad (8a)$$

The integration of Eq. (8a) with respect to $\Phi$ leads to the following expression for $S(\Phi)$, which corresponds to the function for $\tilde{P}_{OSM}$, introduced by Hoehler et al. [51]:

$$S(\Phi)/S_0 = 1.097 - 2.433(2 + 0.5476/\Phi)\sqrt{1-\Phi} + 5.869\,\text{arctanh}\sqrt{1-\Phi} \qquad 0.74 < \Phi < 0.99 \tag{8b}$$

Note that Eq. (8a) gives very similar numerical results to those calculated by the model of Kraynik & Reinelt (see Figures 14,15 in Ref. [1]). Therefore, the results presented below, which are based on Eqs. (8a) and (8b), are representative for the models of $\tilde{P}_{OSM}$ both in Ref. [51] and by Kraynik & Reinelt.

Concluding, for given $R_0$ and $\Phi$, we calculate $\tilde{P}_{OSM}$ and $S(\Phi)$ from Eqs.(7)-(8) which are afterwards used to determine $P_C$, $f(\Phi)$, and $R_{FS}$ from Eqs. (4)-(6). The value of $l_S$ is determined from Eq. (2).

**2.2. Geometrical characteristics of bubbles in steadily sheared foam.**

Now we consider two neighboring bubbles, sliding along each other in sheared foam, see Fig. 1. The bubbles are placed in two neighboring planes of the assumed fcc-structure and these planes move with respect to each other, as a result of applied shear stress. The coordinate system is attached to the geometrical center of one of the bubbles, whereas the other bubble is assumed to move with constant velocity, $u$, along the $x$-axis. The angle formed between $x$-axis and the line connecting bubble centers is denoted by $\alpha(t)$, see Fig. 1.

The relative position of the bubbles can be described by the distance between their geometrical centers:

$$l(t) = \left[l_m^2 + (ut - x_0)^2\right]^{1/2} \qquad l_m = l_S\sqrt{3}/2 \tag{9}$$

where $l_m$ is the minimal distance between the geometrical centers of the bubbles, realized at $\alpha = \pi/2$, see Fig. 1. The coordinate $x_0 = (l_0^2 - l_m^2)^{1/2}$ in Eq. (9) corresponds to the moment of film formation between the bubbles and $l_0$ is the respective center-to-center distance. The dependence $\alpha(t)$ can be also found from Eq. (9), by using the relation, $\alpha(t) = \arcsin[l_m/l(t)]$.

The radius of the planar film between the bubbles, $R_F(t)$, is found from geometrical consideration, by assuming that the radius of curvature of the imaginary "single-film bubble", $R_{\text{eff}}$, is the same for the static and sheared foams (see Fig. 2):

$$R_F(t) = \left[ R_{EFF}^2 - l(t)^2 / 4 \right]^{1/2} \tag{10}$$

In this way we can determine from Eqs.(9)-(10) the relative position of the bubbles $l(t)$, the respective angle $\alpha(t)$, and the film radius, $R_F(t)$, as a function of time, with $t = 0$ being the moment of film formation. To complete the description of bubble dynamics, we should define the moment of film formation, $t = 0$, and determine the dependence of film thickness on time, $h(t)$. The latter two tasks are solved in the following sections 2.3 and 2.4.

**2.3. Description of the liquid flow in the foam film.**

For description of liquid flow in the planar film formed between two sliding bubbles, we use the lubrication equation. Here the foam film is considered in cylindrical $rz\varphi$-coordinate system, whose origin is located in the film center and $z$-axis is perpendicular to the film plane, see Fig. 2(b). In this coordinate system, the lubrication equation reads:

$$\frac{\partial P}{\partial r} = \mu \frac{\partial^2 V_r(r, \varphi, z, t)}{\partial z^2} \qquad (r\text{-component; 11a})$$

$$\frac{\partial P}{r \partial \varphi} = \mu \frac{\partial^2 V_\varphi(r, \varphi, z, t)}{\partial z^2} \qquad (\varphi\text{-component; 11b})$$

$$\frac{\partial P}{\partial z} = 0 \qquad (z\text{-component; 11c})$$

where $\mu$ is liquid viscosity and $P(r,\varphi,t)$ is local pressure in the liquid layer between the bubbles. In lubrication approximation, $P$ does not depend on $z$, whereas the radial and angular components of the fluid velocity, $V_r(r,\varphi,z,t)$ and $V_\varphi(r,\varphi,z,t)$, are functions of all three space-coordinates and time. Note that in the lubrication approximation $V_z$ is much smaller in magnitude than $V_r$ and $V_\varphi$ and, for this reason, it is not considered explicitly, except as a boundary condition for the continuity equation, Eq. (15).

In the following consideration, lubrication equation (11a)-(11c) is solved under the assumption that bubble (drop) surfaces are tangentially immobile, i.e. with non-slip boundary condition for the liquid at the bubble (drop) surface. This tangential immobility of the surfaces might be due to high viscosity of the fluid inside emulsion drops (in the case of sheared emulsion) and/or to high Marangoni stress on the drop/bubble surface, created by adsorbed surfactant/polymer molecules [52-58]. The respective boundary conditions for the fluid velocity at the surfaces of the planar film read:

$$V_r(r, \varphi, z = \pm h/2, t) = \mp (u/2) \sin \alpha(t) \cos \varphi \tag{12a}$$

$$V_\varphi(r,\varphi,z=\pm h/2,t) = \pm(u/2)\sin\alpha(t)\sin\varphi \tag{12b}$$

where $h(t)$ is the instantaneous film thickness, and $u\sin\alpha(t)$ is the projection of the relative bubble velocity in the plane of the film.

The integration of Eqs. (11a)-(11b), along with boundary conditions (12a)-(12b), leads to:

$$V_r(r,\varphi,z,t) = \frac{1}{2\mu}\frac{\partial P}{\partial r}\left(z^2 - \frac{h^2}{4}\right) - (uz/h)\sin\alpha(t)\cos\varphi \tag{13a}$$

$$V_\varphi(r,\varphi,z,t) = \frac{1}{2\mu r}\frac{\partial P}{\partial \varphi}\left(z^2 - \frac{h^2}{4}\right) + (uz/h)\sin\alpha(t)\sin\varphi \tag{13b}$$

To determine the dynamic pressure inside the film, we use the continuity equation and the boundary conditions for the normal component of velocity, $V_z$:

$$\frac{1}{r}\frac{\partial(rV_r)}{\partial r} + \frac{\partial V_\varphi}{r\partial\varphi} + \frac{\partial V_z}{\partial z} = 0 \tag{14}$$

$$V_z(z=\pm h/2) = \pm\frac{1}{2}\frac{dh}{dt} \tag{15}$$

The integration of Eq. (14) from $z = -h/2$ to $z = h/2$ leads to the following differential equation for $P(r,\varphi)$:

$$r^2\frac{\partial^2 P}{\partial r^2} + r\frac{\partial P}{\partial r} + \frac{\partial^2 P}{\partial\varphi^2} = \left(\frac{12\mu r^2}{h^3}\right)\frac{dh}{dt} \tag{16}$$

Equation (16) is solved under boundary condition:

$$P(r=R_F) = P_0 \tag{17}$$

which implies that the pressure of the liquid at the film periphery is not affected by the viscous friction in the film. By following standard approach, one can solve Eq. (16) to obtain the following expression:

$$P(r) = P_0 + \frac{3\mu}{h^3}\frac{dh}{dt}(r^2 - R_F^2) \tag{18}$$

which shows that the dynamic pressure in the film does not depend on the angular coordinate, φ. The latter result is a direct consequence of the assumed boundary condition, Eq.(17), which is independent of φ.

The introduction of $P(r)$ from Eq. (18) into Eqs. (13a)-(13b) leads to:

$$V_r(r,\varphi,z,t) = \frac{3r}{h^3}\frac{dh}{dt}\left(z^2 - \frac{h^2}{4}\right) - (uz/h)\sin\alpha(t)\cos\varphi \tag{19a}$$

$$V_\varphi(\varphi,z,t) = (uz/h)\sin\alpha(t)\sin\varphi \tag{19b}$$

Note that $V_\varphi$-component of the fluid velocity does not depend on the radial $r$-coordinate.

Next, to determine the velocity of film thinning, $dh/dt$, we use the normal force balance [54]:

$$2\pi\int_0^{R_F} P_d(r)r\,dr = \pi R_F^2\left[P_C - \Pi(h)\right] \tag{20}$$

which implies that the dynamic pressure inside the film, $P_d(r) = [P(r) - P_0]$, acting on the film surface from the film interior, is counterbalanced by the difference between the capillary pressure of the bubble, $P_C$, and the disjoining pressure, $\Pi(h)$. As usual, $\Pi(h)$ accounts for the action of the surface forces (per unit area of the film), which could include electrostatic, van der Waals, steric and other types of forces [55,59,60]. Introducing Eq. (18) into Eq. (20) and performing the integration, we obtain the well-known Stefan-Reynolds equation for the velocity of film thinning [52,54]:

$$\frac{dh}{dt} = -\frac{2\left[P_C - \Pi(h)\right]h^3}{3\mu R_F^2} \tag{21}$$

Introducing Eq. (21) into Eq. (19a), we derive the following expression for the radial component of the fluid velocity in the film, $V_r$:

$$V_r(r,\varphi,z,t) = -\frac{2r}{\mu R_F^2}(P_C - \Pi(h))\left(z^2 - \frac{h^2}{4}\right) - (uz/h)\sin\alpha(t)\cos\varphi \tag{22}$$

Thus we derived expressions for the two components of the fluid velocity in the film, Eqs. (19b) and (22), which contain the unknown function $h(t)$. To determine the dependence $h(t)$ we integrate numerically Eq. (21) with the initial condition $h(t=0) = h_0$:

$$\int_{h_0}^{h} \frac{dh}{\left[P_C - \Pi(h)\right]h^3} = -\frac{2}{3\mu}\int_0^t \frac{dt}{\left[R_F(t)\right]^2} \qquad (23)$$

Equation (23) is used in the numerical calculations, discussed in section 4.2 below. Explicit expression for the initial film thickness, $h_0$, is given by Eq. (29) below.

If one neglects the surface forces in Eq. (23), assuming $\Pi(h) \ll P_C$, and using Eq. (10) for the dependence $R_F(t)$, one obtains the following explicit expression for the film thickness $h(t)$:

$$\frac{1}{h^2} = \frac{1}{h_0^2} + \frac{16 P_C}{3\mu} \frac{1}{u\sqrt{4R_{EFF}^2 - l_m^2}} \left[ \text{ArcTanh}\left(\frac{\sqrt{l_0^2 - l_m^2}}{\sqrt{4R_{EFF}^2 - l_m^2}}\right) + \text{ArcTanh}\left(\frac{ut - \sqrt{l_0^2 - l_m^2}}{\sqrt{4R_{EFF}^2 - l_m^2}}\right) \right] \qquad (24)$$

at negligible contribution of $\Pi(h)$

Therefore, at negligible $\Pi(h)$, one can introduce $h(t)$ from Eq. (24) into Eqs. (19b) and (22) to derive explicit expressions for $V_r$ and $V_\varphi$.

### 2.4. Formation of the foam film between colliding bubbles.

To complete our calculation scheme, we should specify the initial moment of formation of the planar film between the colliding bubbles and, thus, to determine the quantities $h_0$, $R_{F0}$, $l_0$, etc. For this purpose, the hydrodynamic approach of Ivanov and co-authors is applied [56-58]. To illustrate the physical meaning of the main steps in this approach and to explain the modifications needed to describe our system, we discuss first the case of negligible contribution of the surface forces, $\Pi(h) \ll P_C$.

Ivanov et al. [56,58] showed theoretically that, if two bubbles collide along their centerline under the action of an external force, $F$, the bubble surfaces flatten and a planar film forms, when the dynamic pressure in the center of the contact zone, $P_d(r=0)$, becomes equal to the bubble capillary pressure, $P_C$. The detailed hydrodynamic calculations of $P_d(r)$, based on the lubrication model for the friction between the surfaces of the colliding bubbles, show that the initial thickness of the formed film depends only on the driving force, $F$, and on the interfacial tension of the bubbles/drops, $\sigma$ [56]

$$h_0 = \frac{F}{2\pi\sigma} \qquad (25)$$

In our system, the bubbles approach each other at given relative velocity, $u$, which is determined by the shear rate in the foam. To adapt the approach from Ref. [56] to our case, we use the fact that the Reynolds number in the gap between the bubbles is low, $Re_h \equiv uh/\mu$

<< 1, so that the inertial and convective terms in the Navier-Stokes equation are negligible and the flow in the gap can be considered as quasi-steady (similar assumptions are used to derive the Stefan-Reynolds equation, Eq. (21)). Thus we can assume that the external force in Eq. (25) is equal (just before the film formation) to the hydrodynamic resistance force between the non-deformed bubbles, which in turn can be estimated from Taylor's formula [55]:

$$F = 3\pi\mu R_N^2 u_z / 2h \qquad (26)$$

Here $R_N$ is the radius of curvature of the approaching bubble surfaces in the zone of their contact, and $u_z(t) = u\cos\alpha(t)$ is the velocity component along the line connecting bubble centers, see Fig. 2(b). It is reasonable to assume that the mean curvature of the colliding bubbles in the contact zone is approximately equal to the mean curvature of the nodes in the static foam, which can be expressed through the capillary pressure of the deformed bubbles, $P_C(\Phi)$. In other words, we assume $R_N \approx 2\sigma/P_C$ and use Eq. (6) to calculate $P_C$.

Combining Eqs. (25) and (26), we derive the following expression for the inversion film thickness in sheared foam, at negligible contribution of the surface forces:

$$h_0 \approx \left[(3/4)\cos\alpha_0 \tilde{u}\right]^{1/2} R_N \qquad (27)$$

where $\alpha_0 = \alpha(t=0)$ is the angle in the moment of film formation and $\tilde{u} = (\mu u/\sigma)$ is the dimensionless relative velocity of the two neighboring planes of bubbles in the sheared foam. Note that $\tilde{u}$ is different from the conventional capillary number for foam and emulsion shear, $Ca = \mu\dot{\gamma}R_0/\sigma$, see section 2.5 below.

Following the approach from Ref. [56], we can estimate also the initial radius of the formed planar film, $R_{F0}$. For this purpose one assumes that immediately after film formation, the pushing force, $F$, becomes equal to the hydrodynamic force resisting the planar film thinning (thus neglecting the friction in the meniscus region surrounding the film in comparison with the friction inside the film, see Section 3 for further discussion). On its turn, this hydrodynamic resistance force in the film can be estimated by multiplying the average hydrodynamic pressure in the film (equal to $P_C$) by the film area, $F \approx P_C\left(\pi R_{F0}^2\right)$ [54]. Thus, by using Eqs. (25)-(26) one derives the following equation for the initial film radius [56]:

$$R_{F0} = (h_0 R_N)^{1/2} \qquad (28)$$

Next, we determine the initial coordinate of the moving bubble in the moment of film formation, $l_0$, by setting $R_F = R_{F0}$ from Eq. (28) into Eq. (10), thus completing the st of equations needed to describe the film dynamics at negligible $\Pi(h)$.

In a recent paper [58], Ivanov and co-authors extended their approach to account for the effect of surface forces on $h_0$. In this case, the pressure balance inside the gap between the colliding bubbles (including the disjoining pressure) leads to the following transcendental equation for the initial film thickness (adapted to our case from Eq. (16) in [58]):

$$h_0 = \frac{R_N \left[ (3/4)\cos\alpha_0 \tilde{u} \right]^{1/2}}{\left[ 1 - \Pi(h_0)/P_C \right]} \qquad (29)$$

where $\Pi(h_0)$ is the disjoining pressure of a film with thickness $h_0$. As seen from Eq. (29), attractive surface forces (e.g., van der Waals attraction), which correspond to negative values of $\Pi(h)$, lead to smaller initial film thickness, $h_0$, whereas repulsive surface forces (e.g., electrostatic repulsion) lead to larger $h_0$, as compared to that for negligible surface forces. For detailed analysis of these relations, see Ref. [58].

Following the approach from Ref. [54], one can show that the initial film radius, $R_{F0}$, at significant disjoining pressure, can be estimated by an expression, which is derived after substituting $h_0$ from Eq. (29) into Eq. (28):

$$R_{F0} = \frac{R_N \left[ (3/4)\cos\alpha_0 \tilde{u} \right]^{1/4}}{\left[ 1 - \Pi(h_0)/P_C \right]^{1/2}} \qquad (30)$$

As in the case of negligible $\Pi(h)$, the bubble coordinate in the moment of film formation, $l_0$, is found by setting $R_F = R_{F0}$ in Eq. (10).

**2.5. Energy of viscous dissipation inside the foam film and average viscous stress in sheared foam.**

In this section we calculate first the energy dissipated in one foam film (between two bubbles) from the moment of film formation to the moment of its disappearance. Afterwards, this energy is used to calculate the average viscous stress in the foam.

The rate of energy dissipation inside the film is calculated by the equation [61]:

$$-\frac{dE_{DF}}{dt} = \mu \int_0^{2\pi} \int_0^{R_F} \int_{-h/2}^{h/2} \left[ \left( r\frac{\partial}{\partial r}\left(\frac{V_\varphi}{r}\right) + \frac{1}{r}\frac{\partial V_r}{\partial \varphi} \right)^2 + \left( \frac{\partial V_\varphi}{\partial z} \right)^2 + \left( \frac{\partial V_r}{\partial z} \right)^2 \right] r\, dz\, dr\, d\varphi \qquad (31)$$

where the subscript "DF" means "energy dissipation inside the film". After substituting the partial derivatives of $V_r$ and $V_\varphi$ from Eqs. (22) and (19b) into Eq. (31) and integrating, one obtains:

$$-\frac{dE_{DF}}{dt} = \pi\mu(uR_F\sin\alpha)^2 / h + \frac{2\pi}{3\mu}[P_C - \Pi(h)]^2 h^3 \qquad (32)$$

The first term in the right-hand-side of Eq. (32) accounts for the viscous dissipation resulting from the sliding of bubbles with respect to each other (due to foam shear), whereas the second term accounts for the viscous dissipation resulting from film thinning. The numerical calculations showed that, typically, the second term is much smaller than the first one and could be neglected in Eq. (32).

To determine the total energy, $E_{DF}$, which is dissipated inside one film during its existence, we integrate Eq. (32) over the contact time, $t_C$, of the two bubbles:

$$E_{DF} = -\int_0^{t_C} \left\{ \pi\mu(uR_F\sin\alpha)^2 / h + \frac{2\pi}{3\mu}[P_C - \Pi(h)]^2 h^3 \right\} dt \qquad (33)$$

where $R_F$, $\alpha$, $h$, and $\Pi$ are functions of $t$.

The contact time, $t_C$, denotes the period of film existence. The initial moment of film formation, $t = 0$, is defined as explained in section 2.4. For the moment of film disappearance we assume that $R_F$ acquires a sufficiently small value, e.g. $\approx 0.1 R_{F0}$. The numerical calculations showed that the final results for the dissipated energy and viscous friction are practically independent of the particular choice of the final film radius, $R_{FE}$, if the latter is chosen sufficiently small (the choice $R_{FE} = 0$ is inconvenient for numerical calculations, because some of the integrated functions diverge, whereas the final integrals are convergent).

From $E_{DF}$ we can calculate the time-averaged energy dissipation rate per unit foam volume, $\langle \dot{E} \rangle$, which is equal to the macroscopic viscous stress, $\tau_{VF}$, multiplied by the shear rate, $\dot{\gamma}$. The shear rate, $\dot{\gamma} = u/m$, and the conventional capillary number $Ca \equiv (\mu\dot{\gamma}R_0/\sigma) = \tilde{u}R_0/m \approx 0.676\tilde{u}\Phi^{1/3}$ are proportional to the relative velocity of the bubble planes, $u$; here $m(\Phi) = (2/3)^{1/2} l_S \approx 1.479 R_0/\Phi^{1/3}$ is the distance between the planes, Fig. 1, and $\tilde{u} = \mu u/\sigma$ is the dimensionless velocity. To determine $\langle \dot{E} \rangle$, we consider the motion of the bubble plane as a sequence of equivalent steps with length $l_S$. During one such step, the six contacts of an arbitrary chosen "central" bubble with its neighbors, in the "top" and "bottom" planes, undergo partial cycles of type "film formation-thinning-disappearance", like those expressed by Eq. (33). Geometrical consideration shows that these partial cycles could be summed up to 4 equivalent full friction cycles. Thus, for the viscous stress of the foam one derives:

$$\tau_{VF}\dot{\gamma} = \langle \dot{E} \rangle = \frac{1}{2}\left[ 4E_{DF} / (V_B/\Phi)(l_S/u) \right] = 2E_{DF}\Phi u / V_B l_S \qquad (34)$$

where the multiplier 1/2 accounts for the sharing of the dissipated energy inside one film by two neighboring bubbles, $V_B/\Phi$ is the volume occupied by one bubble in the foam, and $l_S/u$ is the duration of one step with length $l_S$. The subscript "VF" reminds that only the viscous friction inside the foam films is considered here.

For the numerical calculations it is convenient to introduce dimensionless quantities $\tilde{E}_{DF} \equiv E_{DF}/(R_0^2 \sigma \tilde{u}^{1/2})$; $\xi_F = R_F/R_0$; $\tilde{R}_N = R_N/R_0$; $\tilde{t} = tu/R_0$; $\tilde{\Pi} = \Pi(h)/P_C$; $\tilde{l}_m = l_m/R_0$; $\tilde{R}_{EFF} = R_{EFF}/R_0$; and $\eta = h/R_0$ (note that, for convenience, different scaling was used for $h$ in Ref. [37]). The dimensionless dissipated energy, viscous stress and effective viscosity are thus presented as:

$$\tilde{E}_{DF} = -\int_0^{\tilde{t}_C} \left( \pi \left[ \sin\alpha(\tilde{t}) \right]^2 \tilde{u}^{1/2} \frac{\left[\xi_F(\tilde{t})\right]^2}{\eta(\tilde{t})} + \frac{8\pi}{3} \frac{1}{\tilde{u}^{3/2}} \left(1-\tilde{\Pi}\right)^2 \frac{\left[\eta(\tilde{t})\right]^3}{\tilde{R}_N^2} \right) d\tilde{t} \quad (35)$$

$$\tilde{\tau}_{VF} = \tau_{VF}/(\sigma/R_0) = 0.39\tilde{u}^{1/2}\Phi\tilde{E}_{DF} \approx 0.474 Ca^{1/2} \Phi^{5/6} \tilde{E}_{DF} \quad (36)$$

$$\tilde{\mu}_{EF} = \mu_{EF}/\mu = \tau_{VF}/\mu\dot{\gamma} = \tilde{\tau}_{VF}/Ca \approx 0.474\Phi^{5/6}Ca^{-1/2}\tilde{E}_{DF} \quad (37)$$

Equations (36) and (37) show that the viscous stress is approximately proportional to $Ca^{1/2}$, whereas the effective viscosity is proportional to $Ca^{-1/2}$, as observed experimentally by Princen and Kiss [1,4]. This scaling is ultimately related to the dependence $h \propto Ca^{1/2}$, predicted by the Stefan-Reynolds equation at negligible $\Pi(h)$, see Eqs. (21) and (24).

Note that $\tilde{E}_{DF}$ is a strong function of bubble volume fraction, $\Phi$, and, therefore, the dependences of $\tilde{\tau}_{VF}$ and $\tilde{\mu}_{EF}$ on $\Phi$ are discussed after presenting the numerical results for $\tilde{E}_{DF}(\Phi)$ in section 4. Equations (36) and (37) are applicable to both emulsions and foams, provided that the surfaces of the drops/bubbles are tangentially immobile and the predominant energy dissipation occurs in the foam films (see sections 3, 4.2 and 5 below for other possibilities).

### 3. Viscous dissipation inside meniscus region.

One relevant question is how significant could be the viscous friction in the curved meniscus regions, surrounding the films, for the processes under consideration. Our theoretical analysis showed that the friction in the meniscus region affects the system in two interrelated ways: (1) additional energy is dissipated in the meniscus region around the films, and (2) the film thinning is affected by this additional friction, so that the dependence $h(t)$ is different. In the current section we upgrade our model from section 2 to account for these additional effects.

For simplicity, we consider the meniscus region as part of a spherical surface with radius of curvature, $R_N \approx 2\sigma/P_C$. Close to the film, this spherical surface could be approximated by parabola, which leads to the following description of the liquid layer thickness in the film and its close neighborhood:

$$H(r,t) = h(t) + \frac{(r - R_F(t))^2}{R_N} \qquad r \geq R_F$$
$$H(r,t) = h(t) \qquad r \leq R_F \tag{38}$$

where $R_F(t)$ is the film radius, see Fig. 3.

For determination of the dynamic pressure in the liquid layer between the bubbles (including the meniscus region) we should modify the equations in section 2.3, by accounting for the dependence of the layer thickness on the radial coordinate, $H(r)$. The boundary condition for the normal velocity component, $V_Z$, in this case is:

$$V_z(z = \pm H(r,t)/2) = \pm \frac{1}{2} \frac{\partial H(r,t)}{\partial t} \tag{39}$$

Integration of Eq. (14) from $z = -H/2$ to $z = H/2$, along with the boundary condition, Eq. (39), leads to the following equation for $P(r,t)$:

$$\frac{\partial P}{\partial r} = \frac{12\mu}{[H(r,t)]^3} \left( \frac{r}{2} \frac{dh}{dt} - \frac{(r - R_F)^2 (2r + R_F)}{3rR_N} \frac{dR_F}{dt} \right) \tag{40}$$

Equation (40) is solved under the boundary condition (an analog of Eq. (17)):

$$P(r = R_F + R_N) = P_0 \tag{41}$$

which implies that the pressure in the liquid layer far away from the planar film, is not affected by the viscous friction. By following standard approach, one can solve Eq. (40) and derive the following expressions:

$$P(r) = P_0 + 6\mu \frac{dh}{dt} \left\{ \int_{R_F}^{r} \frac{rdr}{[H(r,t)]^3} - \int_{R_F}^{R_F+R_N} \frac{rdr}{[H(r,t)]^3} \right\}$$

$$-\frac{4\mu}{R_N}\frac{dR_F}{dt} \left\{ \int_{R_F}^{r} \frac{(r-R_F)^2 (2r+R_F) dr}{r[H(r,t)]^3} - \int_{R_F}^{R_F+R_N} \frac{(r-R_F)^2 (2r+R_F) dr}{r[H(r,t)]^3} \right\}$$

at $r > R_F$ (42a)

$$P(r) = P_0 + \frac{3\mu}{h^3}(r^2 - R_F^2)\frac{dh}{dt} - 6\mu \frac{dh}{dt} \int_{R_F}^{R_F+R_N} \frac{rdr}{[H(r,t)]^3}$$

$$-\frac{4\mu}{R_N}\frac{dR_F}{dt} \int_{R_F}^{R_F+R_N} \frac{(r-R_F)^2 (2r+R_F) dr}{r[H(r,t)]^3}$$

at $r < R_F$ (42b)

which is used in the normal force balance, Eq. (44).

The introduction of $\partial P/\partial r$ from Eq. (40) into Eqs. (13a) leads to:

$$V_r(r,\varphi,z,t) = \frac{6}{[H(r,t)]^3} \left( \frac{r}{2}\frac{dh}{dt} - \frac{(r-R_F)^2(2r+R_F)}{3rR_N}\frac{dR_F}{dt} \right) \left( z^2 - \frac{[H(r,t)]^2}{4} \right) - \frac{uz\sin\alpha(t)\cos\varphi}{H((r,t))}$$

(43a)

$$V_\varphi(r,\varphi,z,t) = \frac{uz}{H(r,t)}\sin\alpha(t)\sin\varphi$$ (43b)

Note that the angular $V_\varphi$-component of velocity depends on the radial $r$-coordinate when the friction inside the Plateau border (PB) is taken into account.

Next, to determine the velocity of film thinning, $dh/dt$, we use the normal force balance expressed in the form:

$$2\pi \int_0^{R_F+R_N} rP_d(r)dr = \pi R_F^2 \left[ P_C - \Pi(h) \right]$$ (44)

where, for simplicity, the surface forces in the curved meniscus region, $\Pi(H>h)$, are neglected (see Refs. [62-63] for discussion and calculations of this effect, which could be important for emulsion drops). In the calculations, we solve numerically Eq. (44) to determine the velocity of film thinning, $h(t)$, which is afterwards introduced in the expressions for $V_r$ and $V_\varphi$, Eqs. (43a) and (43b), to determine the rate of energy dissipation and the viscous stress.

The rate of energy dissipation in the meniscus region is given by the expression:

$$-\frac{dE_{DM}}{dt} = \mu \int_0^{2\pi} \int_{R_F}^{R_F+R_N} \int_{-H/2}^{H/2} \left[ \left( r \frac{\partial}{\partial r}\left(\frac{V_\varphi}{r}\right) + \frac{1}{r}\frac{\partial V_r}{\partial \varphi} \right)^2 + \left(\frac{\partial V_\varphi}{\partial z}\right)^2 + \left(\frac{\partial V_r}{\partial z}\right)^2 \right] r\, dz\, dr\, d\varphi \qquad (45)$$

where the subscript "DM" indicates that the energy dissipation inside the meniscus region (which can be considered as part of the Plateau border) is considered. Note that the rate of energy dissipation inside the planar film is described again by Eq. (32), but keeping in mind that the dependence $h(t)$ is affected by the friction in meniscus region and should be found by solving Eq. (44). After substituting the partial derivatives of $V_r$ and $V_\varphi$ from Eqs. (43a) and (43b) into Eq. (45) one obtains:

$$-\frac{dE_{DM}}{dt} = 2\pi\mu \left[ (u\sin\alpha)^2 \left( \int_{R_F}^{R_F+R_N} \frac{r}{H(r,t)}\left[1+\frac{(r-R_F)^2}{6R_N^2}\right] dr \right) + 12 \int_{R_F}^{R_F+R_N} \frac{1}{r[H(r,t)]^3}\left( \frac{r^2}{2}\frac{dh}{dt} - \frac{(r-R_F)^2(2r+R_F)}{3R_N}\frac{dR_F}{dt} \right)^2 dr \right] \qquad (46)$$

The first term in the right-hand-side of Eq. (46) accounts for the viscous dissipation resulting from the sliding of the bubbles with respect to each other (due to foam shear), whereas the second term accounts for the viscous dissipation resulting from thinning of the meniscus region.

The total viscous stress of the foam is given by Eq. (34), after substituting $E_{DF}$ by the sum $(E_{DF} + E_{DM})$ obtained after integration of Eqns. (33) and (41), respectively, over the contact time of the bubbles, $t_C$.

**4. Numerical results.**

**4.1. Negligible friction in the meniscus region and negligible surface forces, $\Pi(h) \ll P_C$.**

We start this section with the simplest possible case, in which both the surface forces and the friction in the meniscus region are negligible. All calculations were performed with both models for $P_{OSM}$, Eqs. (7a)-(7b) and (8), which gave very similar numerical results, except for the range of low volume fractions, $\Phi \to \Phi_{CP}$. To simplify the presentation, only results obtained by using Eq. (8) for $P_{OSM}$ are presented in the figures below, whereas the final interpolation formulae for the viscous stress are presented for both models to allow direct comparison.

By using the model presented in sections 2.3, we calculated the geometrical parameters and the viscous stress for a range of capillary numbers, $10^{-7} \leq Ca \leq 10^{-2}$, and volume fractions of the dispersed bubbles/drops, $0.80 \leq \Phi \leq 0.98$. The numerical results obtained for the film thickness and radius, $h(t)$ and $R_F(t)$, and for the viscous stress, $\tau_V$, in sheared foam/emulsion are discussed below.

The calculations showed that the dimensionless initial film thickness, $\eta_0 = h_0/R_0$, is proportional to $Ca^{1/2}$, that is $h_0/(R_0 Ca^{1/2})$ remains almost constant for fixed value of $\Phi$, see Eq. (27), and decreases approximately twice while increasing $\Phi$ from 0.80 to 0.98. The ratio $h_0/(R_N Ca^{1/2}) \approx 0.8$ is very weakly dependent on both $\Phi$ and $Ca$. Thus we can conclude that the initial film thickness, $h_0$, scales approximately as $Ca^{1/2}$ upon variation of the shear rate, liquid viscosity and surface tension, and with $R_N = 2\sigma/P_C$ upon variation of the bubble size and air volume fraction.

The calculations showed that the dimensionless film radius, $\xi_F = R_F(\tilde{t})/R_0$, passes through a well-defined maximum at $l = l_m$ (corresponding to $\tilde{t} = \tilde{x}_0 \approx 1$) in the process of bubble sliding, see Fig. 4(a). The height of this maximum increases with the increase of $\Phi$, reflecting the more pronounced bubble deformation. The shape of the curves $R_F(\tilde{t})/R_0$ is weakly dependent on $Ca$.

The evolution of the scaled dimensionless film thickness, $h(\tilde{t})/(R_0 Ca^{1/2})$ is plotted in Fig. 4(b), for $Ca = 10^{-4}$ and different volume fractions. One sees that the film thins rapidly during the initial stage of bubble collision, due to the relatively small initial film radius (cf. Fig. 4(a) and Eq. (24)) and large initial thickness. During the following period, when the film radius is relatively large, the film thins much slower and $h(\tilde{t})/(R_0 Ca^{1/2})$ remains almost constant (see the plateau region around $\tilde{t} \approx 1$ in Fig. 4(b)). During the final stage of bubble passage, the film radius decreases rapidly and, as a result, the film thinning accelerates just before the bubble detachment and film disappearance. The effect of $\Phi$ on the dimensionless film thickness, $h(\tilde{t})/(R_0 Ca^{1/2})$, is moderate – the thickness decreases about twice upon increase of $\Phi$ from 0.80 to 0.98, Fig. 4(b). The numerical calculations showed also that the scaled film thickness, $h(\tilde{t})/(R_0 Ca^{1/2})$, is weakly dependent of $Ca$, (i.e. the dimensional thickness is proportional to $Ca^{1/2}$) and that the predominant energy dissipation during the film existence occurs around $\tilde{t} \approx 1$, i.e. when the film radius is the largest.

In Fig. 5 we compare the calculated dimensionless film thickness at maximal film radius (at $l = l_m$, i.e. at $\tilde{t} = \tilde{x}_0 \approx 1$), denoted hereafter as $\eta_1 \equiv h_1/R_0$, for various capillary numbers and air volume fractions. One sees from Fig. 5(a) that $\eta_1$ depends weakly on $Ca$, which means that $h_1 \propto Ca^{1/2}$. Also, the thickness $h_1$ scales with $R_N^{1/2}$ (viz. $P_C^{1/2}$), see Fig. 5(b). Note that the latter scaling stems directly from the Reynolds equation, Eq. (24), in which the first term containing the initial film thickness, $h_0$, is negligible at $\tilde{t} \approx 1$. Therefore, the scaling of $h$ in the region of predominant energy dissipation in the film is determined by the second term in Eq. (24).

From these numerical results we can conclude that the variation of $\Phi$ affects the friction mainly by changing the bubble capillary pressure and film radius, whereas the variation of $Ca$ affects mostly the relative velocity of the bubbles and the film thickness.

Numerical results for the dependence of the dimensionless dissipated energy per film, $\tilde{E}_{DF} = E_{DF}/(R_0^2 \sigma \tilde{u}^{1/2})$, as a function of $Ca$, are shown with symbols in Fig. 6 for five values of $\Phi$. One sees that the dimensionless energy is a very weak function of $Ca$, which shows that the viscous stress scales approximately as $Ca^{1/2}$ (see Eq. (36)). The effect of bubble volume fraction, $\Phi$, on $\tilde{E}_{DF}$ is rather significant. We found that the numerical results in the ranges $0.80 \leq \Phi \leq 0.99$ and $10^{-7} \leq Ca \leq 10^{-2}$ could be described very well by the following empirical functions:

$$\tilde{E}_{DF} \approx 1.7 Ca^{-0.035} / (1-\Phi)^{0.5} \qquad P_{OSM} \text{ from Eq. (7a-7b)} \qquad (47a)$$

$$\tilde{E}_{DF} \approx 2.45 Ca^{-0.03} (\Phi - 0.74)^{0.1} / (1-\Phi)^{0.5} \qquad P_{OSM} \text{ from Eq. (8)} \qquad (47b)$$

Note that both functions exhibit very weak dependence of $\tilde{E}_{DF}$ on $Ca$ and significant dependence on $\Phi$. The comparison of the two functions showed that both give very similar numerical results in the range of volume fractions of main interest, $0.80 \leq \Phi \leq 0.99$ (typically within 4-5 % difference), and significant deviations between the two expressions are observed only when approaching $\Phi_{CP} \approx 0.74$.

Equations (47) were introduced into Eqs. (36) and (37) to obtain our final model expressions for the contribution of film friction into the foam viscous stress and effective viscosity

$$\tilde{\tau}_{VF} \approx 0.806 Ca^{0.465} \Phi^{5/6} / (1-\Phi)^{0.5} \qquad P_{OSM} \text{ from Eq. (7a-7b)} \qquad (48a)$$

$$\tilde{\tau}_{VF} \approx 1.162 Ca^{0.47} \Phi^{5/6} (\Phi - 0.74)^{0.1} / (1-\Phi)^{0.5} \qquad P_{OSM} \text{ from Eq. (8)} \qquad (48b)$$

$$\tilde{\mu}_{EF} \approx 0.806 Ca^{-0.535} \Phi^{5/6} / (1-\Phi)^{0.5} \qquad P_{OSM} \text{ from Eq. (7a-7b)} \qquad (49a)$$

$$\tilde{\mu}_{EF} \approx 1.162 Ca^{-0.53} \Phi^{5/6} (\Phi - 0.74)^{0.1} / (1-\Phi)^{0.5} \qquad P_{OSM} \text{ from Eq. (8)} \qquad (49b)$$

(all at negligible contribution of $\Pi(h)$)

Note that Eqs. (47)-(49) should be used in their range of validity only. The extrapolation to $\Phi \to 1$ is not justified, because the films become very thin at high volume fractions (due to the high capillary pressure of the bubbles) and the surface forces, which were neglected in these calculations, become important in the normal force balance, Eq. (20). Therefore, the upper limit of using these equations is determined by the comparison of the thickness of the dynamic films between the sliding bubbles, $h(t)$, with the extent of surface

forces - typically between 1 and 10 nm, depending on the specific surfactants and electrolytes used (see section 4.2 below).

One can estimate whether the surface forces should be considered, by invoking the fact that in the absence of surface forces $h(\tilde{t} \approx 1) \approx 0.2 R_0 Ca^{1/2}$ - see Fig. 4(b). Thus, for $Ca \approx 10^{-4}$, one estimates $h \approx 2 \times 10^{-3} R_0$, which corresponds to $h \approx 2$ μm for bubbles with $R_0 \approx 1$ mm and to $h \approx 2$ nm for drops with $R_0 \approx 1$ μm. We see that the contribution of the surface forces should be typically negligible in foams, and could be rather important in emulsions, due to the small size of the drops and the resulting thinner films in emulsions. The effect of surface forces on the viscous stress is illustrated with numerical results in the following section 4.2.

The lower boundary of $\Phi$, above which our model is still acceptable, is determined mainly by the model assumption that the bubbles form planar films while sliding along each other. It is difficult to assess theoretically around which volume fraction this assumption would fail, because even for $\Phi < 0.74$ the bubbles deform while passing along each other in sheared foam, due to geometrical constraints and hydrodynamic interactions. The comparison of model predictions with available experimental data, performed in section 5 below, shows that Eqs. (47) and (48) describe satisfactorily the data at least in the range of volume fractions $0.80 \leq \Phi \leq 0.98$.

**4.2. Effect of surface forces on film thickness and on viscous stress.**

As explained in the preceding section, disjoining pressure, $\Pi(h)$, could be important for the studied phenomena, if the thickness of the dynamic films between the sliding drops/bubbles in the sheared emulsion/foam is comparable to the extent of the surface forces. Since this is more typical for emulsions, the estimates and their explanations in the current section are presented for drops in sheared emulsions. However, the same formulas could be applied to sheared foams, after selecting appropriate values for the governing parameters.

To illustrate the main effects of surface forces, we use the DLVO expression for $\Pi(h)$, which accounts for the van der Waals and electrostatic interactions between the drop surfaces [59-60]:

$$\Pi(h) = -\frac{A_H}{6\pi h^3} + 64 n_0 k_B T \left( \tanh \frac{e \Psi_S}{4 k_B T} \right)^2 \exp(-\kappa h) \qquad (50)$$

As usual, $A_H$ denotes Hamaker constant, $n_0$ is electrolyte number concentration, $k_B T$ is thermal energy, $e$ is elementary charge, $\Psi_S$ is electrical surface potential of the drops, and $\kappa$ is inverse Debye screening length.

As seen from Eqs. (20)-(23) and (29)-(30), the surface forces become important for the processes of film formation and thinning, and for the viscous friction, when the dimensionless disjoining pressure, $\widetilde{\Pi} \equiv \Pi(h)/P_C$, becomes comparable to unity. Furthermore, as seen from Eq. (23) the film thinning stops at $\widetilde{\Pi} = 1$, because the latter condition corresponds to films with equilibrium thickness, $h_{EQ}$. In such films, the capillary pressure driving the film thinning is exactly counterbalanced by the repulsive disjoining pressure.

To illustrate the importance of this effect, in Fig. 7 we plot the dimensionless film thickness, $h(\tilde{t})/(R_0 Ca^{1/2})$, as a function of the dimensionless time, $\tilde{t}$, for emulsion drops with different radii, at fixed all other parameters. One sees that, for these particular parameters, the *dimensionless* thickness for the big drops with $R_0 = 100$ μm follows the same curves as those shown in Fig. 4(b) for systems with negligible $\Pi(h)$, whereas the dimensionless thickness is significantly larger for the smaller drops with $R_0 = 10$ and 1 μm. The reason is that for these smaller drops, the thickness of the dynamic films becomes comparable with the range of the electrostatic repulsion and the dynamic films acquire their equilibrium thickness, $h_{EQ}$, during most of the film existence period (note that the actual *dimensional* film thickness is smaller for the smaller drops).

According to Eq. (50), for given material properties of the drops (viz., given values of $A_H$ and $\Psi_S$), the contribution of $\Pi(h)$ on the viscous friction depends on electrolyte concentration (which controls electrostatic repulsion), and on surface tension and drop radius (which determine $P_C$). To illustrate the effect of these factors on the viscous friction in sheared emulsions, we present first the scaled dimensionless film thickness at $l = l_m$, $h_1/(R_N Ca)^{1/2}$, as a function of $Ca$, at different electrolyte concentrations and radii of the drops. As seen from the results shown in Fig. 8(a),(b), when the electrostatic forces are significant, $h_1/(R_N Ca)^{1/2}$ decreases with the increase of $Ca$ (at given $R_0$ and $C_{EL}$) and/or with the increase of $R_0$ and $C_{EL}$ (at given $Ca$). In fact, in all these cases the dimensional thickness remains constant, $h_1 \approx h_{EQ}$, while varying $Ca$ and, as a result, the dimensionless film thickness $h_1/(R_N Ca)^{1/2} \propto Ca^{-1/2}$.

One sees from Eq. (35) that the first term on the right-hand-side is zero when $h = h_{EQ}$ = const during the process of bubble sliding. In this case, the second term could be integrated to derive the following expression for the dissipated energy in the film and for the viscous stress:

$$-\tilde{E}_{DF} = \frac{\pi \tilde{u}^{1/2} \tilde{l}_m^2}{\eta_{EQ}} \left( \frac{\tilde{R}_{eff}^2}{\tilde{l}_m} \left( \arctan \frac{\tilde{t}_c - \tilde{x}_0}{\tilde{l}_m} + \arctan \frac{\tilde{x}_0}{\tilde{l}_m} \right) - \frac{\tilde{t}_c}{4} \right) \qquad (51)$$

$$\tilde{\tau}_V = 0.39 \tilde{u} \Phi \frac{\pi \tilde{l}_m^2}{\eta_{EQ}} \left( \frac{\tilde{R}_{eff}^2}{\tilde{l}_m} \left( \arctan \frac{\tilde{t}_c - \tilde{x}_0}{\tilde{l}_m} + \arctan \frac{\tilde{x}_0}{\tilde{l}_m} \right) - \frac{\tilde{t}_c}{4} \right) \approx C(\Phi) \frac{Ca}{\eta_{EQ}} \qquad (52)$$

Where the dimensionless quantities introduced before Eq. (35) are used. Equation (52) predicts a linear increase of the viscous stress with shear rate, see Fig. 8(c)-8(d). The numerical multiplier $C(\Phi)$ in Eq. (52) varies from 0.4 at $\Phi = 0.80$ to 0.9 at $\Phi = 0.98$. The value of $\eta_{EQ} = h_{EQ}/R_0$ entering these expressions is determined by the balance $\Pi(h_{EQ}) = P_C(\Phi)$, which in turn is governed by the material parameters of the system, such as electrolyte concentration, surface potential and Hamaker constant.

At fixed material parameters, a threshold value of the capillary number exists, $Ca_{TR}$, above which the contribution of $\Pi(h)$ becomes negligible, because the films become thicker than the extent of surface forces, see Fig. 8. From the data shown in Fig. 8 one sees that the threshold capillary number for electrostatic interactions could be estimated by the relation:

$$Ca_{TR} \approx \frac{2 \times 10^{-3}}{R_0 \sqrt{C_{EL}}} \tag{53}$$

where $C_{EL}$ is the electrolyte concentration in mmol/dm$^3$ and $R_0$ is the initial drop radius in micrometers. Therefore, for electrostatically stabilized emulsions, $Ca_{TR}$ depends exclusively on the initial drop radius and electrolyte concentration (provided that the surface potential is not-too-small, $\Psi_S > 25$ mV). Analysis of similar type could be performed for emulsions and foams stabilized by other types of forces, such as steric repulsion, oscillatory structural forces, etc., once the functional dependence $\Pi(h)$ is defined [55].

Thus we see that the viscous stress is a linear function of the capillary number (viz. of shear rate) for emulsions, in which the films are stabilized by repulsive forces and $h \approx h_{EQ}$. At higher capillary numbers, the dynamic films may become thicker than the range of surface forces and the viscous stress becomes proportional to $Ca^{1/2}$, see Fig. 8(c), 8(d). In other words, the model predicts power-law index $n = 1$ at low shear rates in such systems, with possible transition to $n = 1/2$ at higher shear rates. On the other hand, for systems exhibiting shear thinning of the continuous fluid phase (i.e., $\mu$ decreasing with $\dot{\gamma}$), one should expect lower values of the power low index, $n < 1/2$, because the viscous friction inside the films would be lower at high shear rate (as compared to the prediction of the current model). Conceptually, the effect of shear thinning, $\mu(\dot{\gamma})$, could be incorporated in the current model, but this task is not trivial (e.g., the local viscosity in the films could be different from the viscosity of the bulk liquid, if polymers are adsorbed on the film surfaces), requires additional analysis, and is therefore, postponed for a separate study.

Concluding this section we note that our model could explain power-law indexes from $n = 1$ (films with equilibrium thickness), through $n = 1/2$ (dynamic films containing Newtonian liquid), down to $n < 1/2$ (dynamic films containing a shear-thinning liquid). Another possible reason for obtaining power-law index $n < 1/2$ is described in Section 6 below. However, only in the case of $n = 1/2$, the viscous stress depends on well defined parameters ($\Phi$, $R_0$, and $Ca$), which are usually known when making rheological measurements. In the other cases, the viscous stress depends on additional parameters, which are usually unknown - e.g., the functions $\Pi(h)$ and $\mu(\dot{\gamma})$. For this reason, and to avoid the necessity of using unknown parameters when comparing the model with experimental results, we focus the consideration in the following sections 4.3 and 5 mostly on systems with $n \approx 1/2$.

### 4.3. Effect of viscous dissipation inside the meniscus region.

The effect of viscous friction inside the meniscus region around the films could be estimated as explained in section 3. In the current section we compare numerical results for the viscous stress, with and without included friction in the meniscus region, to assess the importance of this effect.

In Fig. 9 we show numerical results for the dimensionless viscous stress, calculated in the range of capillary numbers $10^{-7} \leq Ca \leq 10^{-2}$, at several air volume fractions, $0.80 \leq \Phi \leq 0.98$. One sees that the contribution of the friction in the meniscus region is relatively small at low capillary numbers, $Ca < 10^{-4}$, and at high volume fractions. The effect of meniscus region is very significant only at high capillary numbers, $Ca > 10^{-3}$, and low volume fractions, $\Phi < 0.95$.

Interestingly, we found that the calculated film thickness, $h(t)$, is noticeably larger when the friction in the meniscus region is taken into account. However, at low capillary numbers, the reduced friction in the film (caused by the increased film thickness) is compensated well by the additional contribution to the friction coming from the meniscus region – as a result, the total friction stress is almost the same with and without account for the meniscus region. In contrast, at high capillary numbers and low $\Phi$, the additional friction in the meniscus region is in excess (compared to the reduced friction in the films) and the total viscous stress is higher when taking into account the meniscus region, Fig. 9.

An appropriate for estimates semi-empirical formula was designed, by considering the total viscous stress as a superposition of the friction in foam films, Eq. (48b), with the friction in meniscus region:

$$\tilde{\tau}_V \approx 0.7\, Ca^{0.47}\, \Phi^{5/6} / (1-\Phi)^{0.5} + 8\, Ca^{0.7}\, \Phi^{5/6} / (1-\Phi)^{0.15} \qquad P_{\text{OSM}} \text{ from Eq. (8)} \qquad (54)$$

$$\tilde{\mu}_{EF} \approx 0.5\, Ca^{-0.535}\, \Phi^{5/6} / (1-\Phi)^{0.5} + 6.2\, Ca^{-0.3}\, \Phi^{5/6} / (1-\Phi)^{0.2} \qquad (55)$$

(at negligible contribution of $\Pi(h)$)

The first term accounts for the friction inside the film, whereas the second term is due to the friction inside the meniscus region. At low values of $Ca$ and/or high values of $\Phi$ the first term dominates (i.e., the prevailing friction is in the film), whereas at high $Ca$ and low $\Phi$ the friction in the meniscus could dominate.

### 5. Comparison of model predictions with experimental data.

Let us compare first the model predictions with the experimental data of Princen and Kiss [4], who measured the viscous stress, $\tau_V(Ca)$, of concentrated emulsions with different oil volume fractions and relatively large mean drop size. Since the emulsions in these experiments are polydisperse, the comparison is made by using the mean volume-surface radius, $R_{32}$, as analog to the drop radius, $R_0$, in our model. The comparison showed a reasonably good agreement of Eq. (54) with all experimental data from Ref. [4], without using any adjustable parameter. As illustration, we show in Fig. 10(a) the theoretical curves and the experimental data from Ref. [4] for two of the used emulsions with volume fractions $\Phi = 0.83$ and $\Phi = 0.96$. One sees a very good description of the data by the model. For the other volume fractions and mean drop sizes studied in [4] the agreement is also reasonably good - no difference larger than 15 % was found for the entire set of data (the agreement was within 10 % for most of the experimental data). Very good agreement was found also between model predictions and our own experimental data, obtained with silicone oil-in-water emulsions, stabilized by the nonionic surfactant tridecylether-polyoxyethylene-8 ($C_{13}EO_8$), Fig. 10(b).

Thus, we can conclude from the comparisons shown in Fig. 10 that our model with account for the friction in the film and meniscus regions, Eq. (54), describes well experimental results for emulsions, which exhibit a flow index $n \approx 1/2$. As explained in section 4.2, the comparison of our model with results obeying other values of the flow index, $n \neq 1/2$, would require additional parameters, which are unknown.

Next, we compare model predictions with experimental results for sheared foams. The comparison for foams with $n \approx 1/2$, which were prepared by mixed solutions containing 0.33 wt % anionic surfactant sodium dodecylpolyoxyethylene-2-sulfate (SLES) and 0.17 wt % zwitterionic surfactant cocoamidopropyl betaine (CAPB) [38], showed also very good agreement without adjustable parameters, see Fig. 11, especially when using the model accounting for the film friction only, Eq. (48). The model accounting for the friction in the meniscus region, Eq. (54), predicted higher viscous stress than the experimental data at $Ca > 10^{-4}$. This comparison indicates that the surface of the bubbles in the meniscus region might be tangentially mobile (at least partially) and the dissipation inside the meniscus region is negligible, as compared to the friction in the films, for these foams.

The comparison of the theoretical viscous stress with the experimental data presented in Refs. [26,38], showed that the foams stabilized by another type of surfactant (salts of fatty acids) exhibit $n < 1/2$ and much higher viscous stress in comparison with SLES+CAPB stabilized foams. Obviously, the model described above cannot explain such higher stresses. One specific feature of fatty acid-based surfactants is their very high surface dilatational modulus, $G_S > 100$ mN/m, in comparison with < 5 mN/m for SLES+CAPB mixtures [26,38]. Because the surface modulus $G_S$ contains a significant loss (viscous) component, $G_{LS}$, one can suggest that the viscous dissipation of energy at the surface of the bubbles (which perform a series of small expansions and contractions around the average surface area in sheared foam) should be taken into account when describing the fatty acid-stabilized foams. Therefore, the effect of surface dissipation of energy is considered in the following section 6.

**6. Surface dissipation of energy in sheared foams.**

In this section we show that the higher viscous stress, observed with foams stabilized by salts of alkylcarboxylic acids [26,38], can be explained by the high surface viscosity of these systems.

To estimate approximately the effect of surface dilatational viscosity on the foam viscous stress, $\tau_V(\dot\gamma)$, we represent the consecutive expansions/contractions of the bubble surface in flowing foam with oscillatory-type of deformations of amplitude $\delta S$ and angular frequency $\omega$:

$$S(t) = S_0 + \delta S \cos(\omega t) \tag{56}$$

Next, we estimate the contribution of the surface dissipation of energy to the total density of energy dissipation in the foam. Following a standard procedure [64], we integrated the energy dissipation rate on the bubble surface, $dE_{DS}/dt = G_{LS}\omega(\delta S/S_0)^2 \sin^2\omega t$, over one oscillatory cycle of the bubble area expansion/contraction, $T = 2\pi/\omega$, and obtained the following expression for the energy dissipated at the surface of one bubble, upon formation and disappearance of one foam film:

$$E_{DS} = \pi G_{LS} S_0 (\delta \ln S)^2 \tag{57}$$

Here $\delta \ln S \equiv \delta S/S_0$ denotes the relative amplitude of the bubble area deformation, as a result of the film formation between two sliding bubbles, and $G_{LS}$ is the surface dilatational loss (viscous) modulus.

Following the same reasoning as in the derivation of Eq. (34), we calculate the time-averaged rate of energy dissipation per unit foam volume, $\langle \dot E_{DS} \rangle$, which is equal to the product of the respective macroscopic viscous stress, $\tau_{VS}$, multiplied by $\dot\gamma$, which leads to the following expression:

$$\tau_{VS} = \langle \dot E_{DS} \rangle / \dot\gamma = 4 E_{DS} \Phi / (t_C V_B \dot\gamma) = 12\pi G_{LS} \Phi m (\delta \ln S)^2 / (R_0 l_S) \tag{58}$$

The multiplier 4 in Eq. (58) stands for the number of films formed upon one elementary step with length $l_S$ of the moving neighboring planes of bubbles (see the discussion after Eq. (34)). Taking into account that $m/l_S \approx 0.82$, we derive the following final expression for the dimensionless stress and effective viscosity:

$$\tilde\tau_{VS} \equiv \tau_{VS} R_0/\sigma \approx 9.8\pi (G_{LS}/\sigma) \Phi (\delta \ln S)^2 \tag{59}$$

$$\tilde{\mu}_{EF} \equiv \mu_{EF}/\mu \approx 9.8\pi(G_{LS}/\sigma)\Phi(\delta \ln S)^2 Ca^{-1} \qquad (60)$$

Note that the dependence of $\tau_{VS}$ on $\dot{\gamma}$ appears only through the possible dependences of $G_{LS}$ and $\delta \ln S$ on $\dot{\gamma}$.

To check properly these theoretical predictions, we measured the foam viscous stress and the surface rheological properties for several surfactant systems, particularly chosen to cover a wide range of surface viscosities [38]. Foam viscous stress, $\tau_V(\dot{\gamma})$, was determined by parallel-plates rheometry for shear rates between 0.1 and 150 s$^{-1}$, as explained in Ref. [26,38]. Surface rheological properties of the foaming solutions were characterized by the oscillating drop method (ODM) in the frequency range $\nu = 0.05$-$0.5$ Hz and at relative amplitude of the area oscillations between 0.2 and 1.5 %.

The reference surfactant system was a mixture of 0.33 wt % anionic surfactant SLES and 0.17 wt % zwitterionic surfactant CAPB [38]. The surface dilatational modulus of this system is rather low, $G_{LS} < 3$ mN/m. Therefore, this reference solution is representative for the typical foaming solutions of synthetic surfactants with low surface modulus. As explained in Ref. [38], the addition of 0.02 wt % of Lauric acid (LAc) or Myristic acid (MAc) to this reference solution leads to significant increase of the surface loss modulus, $G_{LS}$, whose value depends on both the frequency and amplitude of oscillations. As illustration, we show in Fig. 12(a) the scaled loss modulus, $G_{LS}/\sigma$, for the three solutions studied, as a function of the amplitude of area oscillations, $\delta S/S_0 \equiv \delta \ln S$ %, at frequency $\nu = 0.2$ Hz. The high surface modulus measured with MAc and LAc-containing solutions was explained in [38] with the formation of two-dimensional crystal of MAc or LAc molecules in the surfactant monolayer, adsorbed at the solution surface.

The viscous stresses, measured with foams generated from the same three surfactant solutions ($\Phi = 0.9$), are compared in Fig. 11. The experimental data for the reference solution, SLES+CAPB, are very well described by Eq. (48), which accounts for the viscous friction in the foam films only and predicts power-law index $n \approx 0.46$. In contrast, the experimental data for the other two systems, containing LAc or MAc, correspond to $n \approx 0.23$ and lay much higher than the predictions of Eq. (48) or (54). To check whether the surface dissipation of energy could explain this "extra" viscous stress, we subtracted the calculated stress related to friction in the film and meniscus regions ($\tau_{VF} + \tau_{VM}$), see Eq. (54), from the total measured viscous stress, $\tau_V$. The remaining stress is certainly related to the specific surface properties of LAc and MAc-containing systems and is, therefore, denoted by $\tau_{VS}$.

By assuming that the prevailing contribution into $\tau_{VS}$ comes from the surface dissipation of energy (other possibilities are mentioned at the end of this section), we normalized $\tau_{VS}$ in accordance with Eq. (59) and plotted in Fig. 12(b) the dependence $(G_{EFF}/\sigma)(\delta \ln S)^2$, calculated from the foam rheology data, on the foam shear rate, $\dot{\gamma}$. Here $G_{EFF}$ denotes the effective surface loss modulus, determined from $\tau_{VS}$ (for convenience, $\delta \ln S$ is expressed in % throughout the following discussion). One sees from Fig. 12(b) that $(G_{EFF}/\sigma)(\delta \ln S)^2$ increases with the shear rate up to 10 s$^{-1}$ and then decreases. This non-homogeneous dependence is in qualitative agreement with our measurements of the surface rheological properties of MAc and LAc-containing systems, which showed a sharp decrease of the surface modulus at high frequencies and amplitudes of oscillations, probably due to destruction of the crystalline adsorption layer under these conditions [38]. Unfortunately, for technical reasons we could not measure the surface rheological properties by ODM at frequency of oscillations > 0.5 Hz. Therefore, we could not make a direct comparison of the surface properties measured by ODM with foam rheological properties at high frequencies. Thus we restrict our further analysis to the region of low shear rates only, $\dot{\gamma} < 1$ s$^{-1}$.

In this low-rate region we can describe well the experimental data in Fig. 12(b) by the empirical fit $(G_{EFF}/\sigma)(\delta \ln S)^2 = A\dot{\gamma}^{0.18}$, where $A \approx 12.6$ for LAc containing foams, $A \approx 21.2$ for MAc containing foams, and $\dot{\gamma}$ is expressed in s$^{-1}$ (see the dashed lines in Fig. 12(b)). Because the values of $\sigma$ are practically equal and the values of $\delta \ln S$ are expected to be similar for the LAc and MAc-containing systems, the higher stress observed with the MAc system in Fig. 12(b) reflects higher values of $G_{EFF}$, which is in qualitative agreement with the data obtained by the ODM method, cf. Fig. 12(a).

Next, to compare quantitatively the results from the ODM experiments with the results from foam rheological measurements, we assume that (1) the frequency of oscillations in the ODM experiments corresponds approximately to the frequency of collisions between the bubbles, $\nu = \omega/2\pi \approx \dot{\gamma}$, and (2) $G_{LS}$ from the ODM experiments should be equal to $G_{EFF}$ from the foam experiments. From the crossing points of the curves drawn in Fig. 12(a), corresponding to $G_{LS}(\nu) = G_{EFF}(\dot{\gamma})$, we could estimate approximately the amplitude $\delta S/S_0$, at which the results from the foam rheometry agree with the results from the surface characterization of the foaming solutions. One sees from Fig. 12(a) that such agreement is obtained at $\delta S/S_0 \approx 1.5 \pm 0.1\%$ for LAc and $\delta S/S_0 \approx 1.1 \pm 0.1\%$ for MAc containing solutions. Very similar values for $\delta S/S_0$ were determined at the other frequencies studied, which shows that the main dependence of the product $G_{EFF}\delta \ln S$ on the shear rate comes from $G_{EFF}$. This conclusion is supported by the ODM measurements, which showed higher values of $G_{LS}$ at higher frequencies of oscillations.

The estimated amplitude of the average surface deformation of the bubbles in sheared foams, $\delta S/S_0 \approx 1\%$ (at $\Phi = 0.90$), seem rather reasonable, taking into account the various approximations used in the model development, and the estimate showing that formation of a single film with radius $R_F(l_m)$ on the surface of an initially spherical bubble, would correspond to $\delta S/S_0 \approx 0.55\%$. Note that the actual bubble shape in sheared foams is rather complex and the surface oscillations created by the formation of several films on the bubble surface could interfere with each other, thus affecting the average value of $\delta S/S_0$.

Let us note at the end of this section that one could envisage additional possible sources of dissipated energy in the sheared foams/emulsions, which were neglected so far. For example, at high surface elastic modulus, one could expect that the bubbles would be forced to rotate under the action of the torque, created by the bubbles in the neighboring (upper and lower) moving planes. Such a bubble rotation would lead to viscous friction and to respective energy dissipation inside the films formed with the neighboring bubbles *in the same plane* – an effect, which has been neglected in the model. It is difficult to quantify this contribution in the viscous stress, because the rotation speed of the bubbles, as well as the thickness of the films between the bubbles arranged inside a given plane are unknown. Another contribution could come from the migrations of bubbles between neighboring planes. Such a process would lead to additional dissipation of energy in the films between the migrating bubble and its neighbors. Without having experimental information about the frequency of such migration events, one cannot evaluate the importance of this process. Therefore, the detailed consideration of these additional contributions is postponed for subsequent studies.

**8. Conclusions.**

In this paper we explain in more detail and extend further our model from Ref. [37] for the viscous friction in steadily sheared foams and concentrated emulsions. Along with the viscous friction inside the foam/emulsion films, formed between the neighboring bubbles and drops [37], three additional effects are considered here: (1) Surface forces acting between the film surfaces, (2) Energy dissipation on the bubble/drop surface due to the perpetual bubble/drop deformation in the sheared foams/emulsions; (3) Friction in the meniscus region around the planar films. The main conclusions from our study could be summarized as follows:

- At negligible surface dissipation and surface forces, the viscous stress is approximately proportional to the capillary number of power $n \approx 1/2$, i.e. $\tau_V \propto Ca^{1/2}$, see Eqs. (48). This prediction is compared with several sets of experimental data, obtained with emulsions and foams by different research groups, and very good agreement is observed without adjustable parameters, Figs. 10 and 11.
- The surface forces are usually insignificant for steadily sheared foams (except for micrometer-sized bubbles and/or at very low shear rates), because the dynamic films in these systems are typically thicker than the range of surface forces. In contrast, for micrometer-sized drops and bubbles, the surface forces could be important, because their range becomes comparable to the dynamic film thickness. If the films between the sliding drops/bubbles rapidly acquire their equilibrium film thickness, $h_{EQ}$, determined by the balance of the capillary and disjoining pressures, $P_C = \Pi(h_{EQ})$, the power-law index $n \to 1$, see Fig. 8.

- Surface dissipation is important for systems with high surface loss (viscous) modulus, $G_{LS} > 20$ mN/m. In these systems, the power-law index $n < 1/2$, because the surface dissipation is a weak function of capillary number, see Eq. (59) and Fig. 11.
- Friction in the meniscus region around the films becomes significant only at both high capillary number, ca. $Ca > 10^{-4}$, and low air volume fraction, Fig. 9. The resulting additional viscous stress could be described by interpolation formula, Eq. (54).

Summarizing, our model predicts power-law indexes $n \approx 1$ for dynamic films having equilibrium thickness, $n \approx 1/2$ for dynamic films whose thickness is described by the Reynolds equation (in both cases at negligible surface dissipation and for films containing Newtonian liquid), and $n < 1/2$ for significant surface dissipation and/or dynamic films containing a shear-thinning liquid. Therefore, a wide range of power-law indexes, $n$, is predicted, depending on the system characteristics, in agreement with the experimental results published in literature.

We note that the different expressions for the osmotic pressure of emulsions and foams, Eqs. (7)-(8), give very similar results when used to calculate the viscous stress in such systems. From practical viewpoint, most convenient is Eq. (8), because it is simple, sufficiently accurate, and applicable in wide range of volume fractions.

The model described in the current paper has two important extensions, which are currently under development:

First, it can be applied to two-dimensional (2D) foams, after straightforward modification to account for the different bubble arrangement and for the possible foam-wall friction, which is typical for many rheological studies of 2D-foams [14,15,23,24,65-68]. The respective calculations are in progress and will be published in a subsequent paper [69].

Second, the foam and emulsion films are known to undergo a spontaneous jump to smaller thickness, under the action of attractive surface forces (such as the van der Waals and depletion forces), at a certain critical film thickness, $h_{CR}$ which is typically between 5 and 30 nm [52,54,58,70-71]. The ultrathin films, formed after such a jump, lead to strong adhesion between the drops/bubbles confining these films. Therefore, the foam and emulsion jamming, observed at low shear rates [20,22-25], might be intimately related to the spontaneous jump in film thickness at $h_{CR}$. Our preliminary estimates showed that the time, required for film thinning down to $h_{CR}$, might be comparable to the contact time of the bubbles/drops in sheared foams and emulsions, thus suggesting that the jamming phenomenon in these systems could be governed by the rate of film thinning, Eq. (21). This hypothesis is under investigation and the results from the quantitative comparison of the model predictions with experimental results will be presented in a subsequent study [72].

**Acknowledgment.** This study was supported by Unilever R&D, Trumbull, CT, USA, and by the COST action P21 "Physics of drops" of the EC. The authors are indebted to Mrs. M. Temelska from Sofia University for the solution characterization by the ODM method.

# Notation

## Capital latin letters

$A_H$ – Hamaker constant

$C_{EL}$ – electrolyte concentration

$Ca = \mu\dot{\gamma}R_0/\sigma$ - capillary number

$E$ – dissipated energy
    $E_{DF}$ – energy dissipated inside the film region, Eq. (35)
    $E_{DM}$ – energy dissipated in the meniscus region, Eq. (46)
    $E_{DS}$ – energy dissipated on bubble surface
    $\langle \dot{E} \rangle$ - time-averaged rate of energy dissipation per unit foam volume, Eq. (34)

$\tilde{E} = E/\sigma R_0^2 \tilde{u}^{1/2}$ – dimensionless dissipated energy

$F$ – external force, section 2.4

$G$ – surface dilatational modulus
    $G_{LS}$ – loss modulus measured in oscillating drop experiments
    $G_{EFF}$ – loss modulus estimated from rheological experiments with bulk foam

$H$ – liquid layer thickness in the film and its close neighborhood, Eq. (38)

$P$ – pressure
    $P_0$ – reference pressure in the continuous phase (away from the film), which is not affected by the viscous friction
    $P_C$ - capillary pressure of the drops/bubbles, Eq. (6)
    $P_d$ – dynamic pressure inside the film, defined as $P_d(r) = [P(r) - P_0]$
    $P_{OSM}$ – osmotic pressure of the concentrated foam/emulsion, Eqns. (7)-(8)

$\tilde{P} = P/(\sigma/R_0)$ – dimensionless pressure

$R$ – radius
    $R_0$ – radius of non-deformed bubble
    $R_{eff}$ – effective radius of curvature of deformed bubbles, Fig. (2)
    $R_{FS}$ – radius of the foam film between two neighboring bubbles in the static foam, Eq. (4)
    $R_F$ - film radius in sheared foam, Eq. (10)
    $R_{F0}$ – film radius in the moment of film formation, section 2.4
    $R_N$ – radius of curvature of the approaching bubble surfaces in the zone of their contact, (assumed to be equal to the mean curvature of the nodes in static foam, $R_N = 2\sigma/P_C$)

$S$ – area
    $S$ – total surface area of the deformed bubble/drop

$S_0 = 4\pi R_0^2$ – surface area of non-deformed bubble

$S_1$ – area of deformed bubble at $\Phi \to 1$;

$S_{FS}$ – area of bubble surface, occupied by films

$V$ – fluid velocity
    $V_r$ – radial component of velocity vector
    $V\varphi$ – angular component of velocity vector
    $Vz$ – z-component of velocity vector

$V_B$ – volume of non-deformed bubble, $V_B = 4\pi R_0^3/3$

$T$ – temperature

## Small latin letters

$e$ – elementary charge

$f(\Phi)$ – fraction of wall occupied by wetting films, Eq. (5)

$h$ – thickness of the planar film
    $h_0$ – thickness at moment of film formation, section 2.4
    $h_1$ – film thickness at $l = l_m$, Fig. 1
    $h_{EQ}$ – equilibrium film thickness

$k$ – consistency in Herschel-Bulkley model, Eq. (1)

$k_B$ – Boltzmann constant

$l$ – distance between the geometrical centers of two neighboring bubbles, Fig. 1
    $l_S$ – in static foam, Eq. (2)
    $l_m$ – minimal distance between the geometrical centers of the bubbles, Fig. 1
    $l_0$ – in the moment of film formation

$\tilde{l} = l/R_0$ - dimensionless distance between bubble centers

$m$ – distance between the neighboring planes in the assumed fcc-foam structure, Fig. 1

$n$ – power-law index

$r$ – radial coordinate in cylindrical $rz\varphi$-coordinate system, Fig. 2(b).

$t$ - time
    $t_C$ – contact time of bubbles

$\tilde{t} = tu/R_0$ - dimensionless time

$u$ – relative velocity of two neighboring planes of bubbles in sheared foam

$\tilde{u} = \mu u/\sigma$ - dimensionless velocity

$x$ – coordinate in the system attached to the geometrical center of one bubble, whereas the neighboring bubble moves along the $x$-axis, Fig. 2(b).

    $x_0$ – bubble coordinate in the moment of film formation

$\tilde{x}_0 = x_0/R_0$ - dimensionless coordinate

$y$ –coordinate in the system attached to the geometrical center of the bubble, Fig. 2(b).

$z$ – coordinate in cylindrical $rz\varphi$-system, whose origin is located in the film center and $z$-axis is perpendicular to the film plane, Fig. 2(b).

**Capital greek letters**

$\Phi$ - volume fraction of the dispersed phase

$\Phi_{CP}$ - volume fraction of closely packed spheres

$\Pi$ - disjoining pressure (surface forces acting per unit area of the film)

$\tilde{\Pi} = \Pi/P_C$ - dimensionless disjoining pressure

$\Psi_S$ - electrical surface potential

**Small greek letters**

$\alpha$ – angle formed between $x$-axis and the line connecting bubble centers, Fig. 1

    $\alpha_0 = \alpha(t=0)$ - angle $\alpha$ in the moment of film formation

$\varphi$ - angular coordinate in the cylindrical $rz\varphi$-coordinate system, whose origin is located in the film center and $z$-axis is perpendicular to the film plane.

$\kappa$ - inverse Debye screening length

$\dot{\gamma}$ - shear rate

$\eta$ - dimensionless film thickness, $\eta = h/R_0$

    $\eta_1$ - dimensionless film thickness at maximal film radius, $R_F(l_m)$

    $\eta_{EQ}$ - dimensionless equilibrium film thickness

$\mu$ - dynamic viscosity of the liquid

    $\mu_{EF}$ – effective viscosity due to the friction inside the film region, Eqs. (37), (55), (60)

$\tilde{\mu}_{EF} = \mu_{EF}/\mu$ - dimensionless effective viscosity

$\tau$ - stress

$\tau_0$ - yield stress

$\tau_V$ - viscous (shear rate-dependent) stress

$\tau_{VF}$ – viscous stress due to the friction inside the film region, Eq. (36)

$\tau_{VM}$ – viscous stress due to the friction in both film and meniscus regions, Eq. (54)

$\tau_{VS}$ – viscous stress due to the surface dissipation, Eq. (58)

$\tilde{\tau} = \tau/(\sigma/R_0)$ - dimensionless stress

$\sigma$ - interfacial tension

$\omega$ - angular frequency of oscillation

$\nu$ - frequency of oscillation in ODM experiments

$\xi_F = R_F/R_0$ – dimensionless film radius

**<u>Abbreviations</u>**

fcc – face-centered cubic

CAPB – cocoamidopropyl betaine

DLVO – Derjaguin-Landau-Verwey-Overbeek

ODM – oscillating drop method

LAc – Lauric acid

MAc – Myristic acid

SLES – dodecylpolyoxyethylene-2-sulfate

# Figure Captions

**FIG. 1.** (a) Illustration of the process of bubble-bubble friction with consecutive images of bubbles passing along each other in sheared foam. The images are taken by side-observation of foam, sheared between the parallel plates of rheometer. (b) Schematic presentation of the relative motion of the neighboring planes of bubbles in sheared foam (with relative velocity $u$), and of the process of film formation and disappearance between two bubbles, sliding along each other. The relative position of the bubbles is characterized by the distance $l(t)$ or by the angle, $\alpha(t)$, whereas $l_m$ denotes the distance of closest approach of the bubble centers at $\alpha = 90°$ (upper line: side view; bottom line: projection onto the plane of bubbles)

**FIG. 2.** Schematic presentation: (a) Two neighboring bubbles in static foam: $R_{eff}$ is the effective bubble radius, $l_S$ is the distance between bubble geometrical centers, $h_{EQ}$ is equilibrium film thickness, and $R_{FS}$ is film radius. (b) Sliding bubbles in sheared foam: $h$ is film thickness, $R_F$ is film radius, $l$ is center-to-center distance, $\alpha$ is running angle (see Fig. 1) and all these are functions of time. The coordinate system $Oxy$ used to describe the bubble relative position is fixed to the center of "immobile" upper bubble, whereas the center of the moving coordinate system $O_1 rz\varphi$, used in the description of film thinning, is located in the center of the planar film between the bubbles.

**FIG. 3.** Schematic presentation of transient foam film, $0 \leq r \leq R_F$, with the adjacent meniscus region $R_F \leq r \leq R_N + R_F$.

**FIG. 4.** Dimensionless (a) film radius, $\xi_F = R_F(\tilde{t})/R_0$, and (b) film thickness, $h(\tilde{t})/(R_0 Ca^{1/2})$, as functions of dimensionless time, $\tilde{t} = tu/R_0$, calculated for $Ca = 10^{-4}$ and different air volume fractions, $\Phi$. Equation (8) is used for $\tilde{P}_{OSM}$.

**FIG. 5.** (a) Dimensionless film thickness, $\eta_1/Ca^{1/2}$, as a function of the capillary number, $Ca$ for different air volume fractions; $\eta_1$ is defined as $\eta(l = l_m)$, which corresponds to $\tilde{t} = \tilde{x}_0 \approx 1$; (b) Scaled dimensionless film thickness, $h_1/(R_N Ca)^{1/2}$, as a function of $Ca$. Equation (8) is used for $\tilde{P}_{OSM}$.

**FIG. 6.** Dimensionless energy dissipated in one film, $\tilde{E}_{DF} = E_{DF}/\left(R_0^2 \sigma \tilde{u}^{1/2}\right)$, as a function of the capillary number, $Ca$, for different air volume fractions, $\Phi$. The symbols are numerical results, whereas the curves show the empirical interpolation, Eq. (48b). Equation (8) is used for $\tilde{P}_{OSM}$.

**FIG. 7.** Dimensionless film thickness as a function of dimensionless time, $\tilde{t} = tu/R_0$, for emulsions with different drop radii, $R_0$. The other parameters used for the calculations are: $A_H = 4 \times 10^{-21}$ J; $\Psi_S = 100$ mV; $\sigma = 5$ mN/m; $T = 298$ K; $C_{EL} = 1$ mM; $Ca = 10^{-4}$ and $\Phi = 0.9$. Equation (8) is used for $\tilde{P}_{OSM}$.

**FIG. 8.** (a,b) Dimensionless film thickness, $h_1/(R_N Ca)^{1/2}$, and (c,d) dimensionless viscous stress, $\tilde{\tau}_{VF} = \tau/(\sigma/R_0)$, as functions of the capillary number, $Ca$. (a,c) different electrolyte concentrations, (b,d) different drop sizes. The parameters used in the calculations are typical for emulsions stabilized by ionic surfactants: $A_H = 4\times10^{-21}$ J, $\Psi_S = 100$ mV, $\sigma = 5$ mN/m, $T = 298$ K, and $\Phi = 0.9$. Equation (8) is used for $\tilde{P}_{OSM}$.

**FIG. 9.** Dimensionless viscous stress vs. capillary number. Solid curves – with account for the viscous friction in the meniscus region; dashed curves – with account for the friction in the films only. Equation (8) is used for $\tilde{P}_{OSM}$.

**FIG. 10.** Comparison of model predictions with experimental results (symbols) for oil-in-water emulsions: (a) Data from Ref. [4] for $\Phi = 0.83$ and 0.96; (b) Our results obtained by parallel-plates rheometry, by using the method described in Ref. [26]. The solid curves in (a) and (b) are drawn according to Eq. (54), whereas the dashed curve in (b) is drawn with account for the dissipation in the film region only (all curves - without adjustable parameter).

**FIG. 11.** Comparison of the theoretical predictions with experimental data for foams formed from surfactant solutions with different surface loss modulus, $G_{LS}$ (see section 6). The symbols are experimental data, the solid curve is drawn according to Eq. (38) (friction only in the foam films, no adjustable parameter).

**FIG. 12.** (a) Comparison of the surface loss modulus, $G_{LS}$, measured by oscillating drop method (symbols and the solid lines, which are drawn to guide the eye) with the effective modulus $G_{EFF}$, determined from the foam viscous stress (dashed curves). From the cross-points one can determine the effective bubble deformation in the sheared foam, $\delta S/S_0$, which provides agreement between the calculated and the measured foam viscous stress, generated by the surface dissipation of energy, $\tau_{VS}$. (b) Effective surface loss modulus, $G_{EFF}$, determined from the foam viscous stress, after subtracting the contribution from the friction in the foam films and meniscus region, see Eq. (59). The lines are linear interpolations in the region of low shear rates, $\dot{\gamma} \leq 1 s^{-1}$.

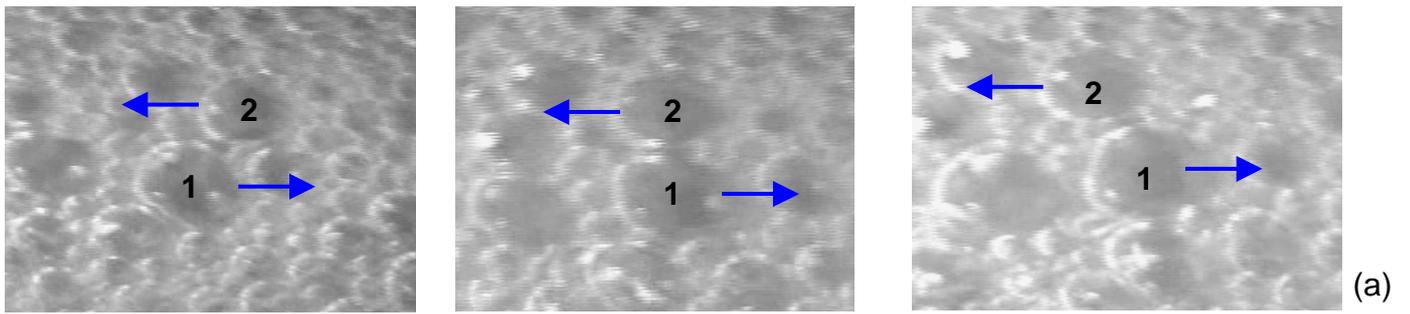

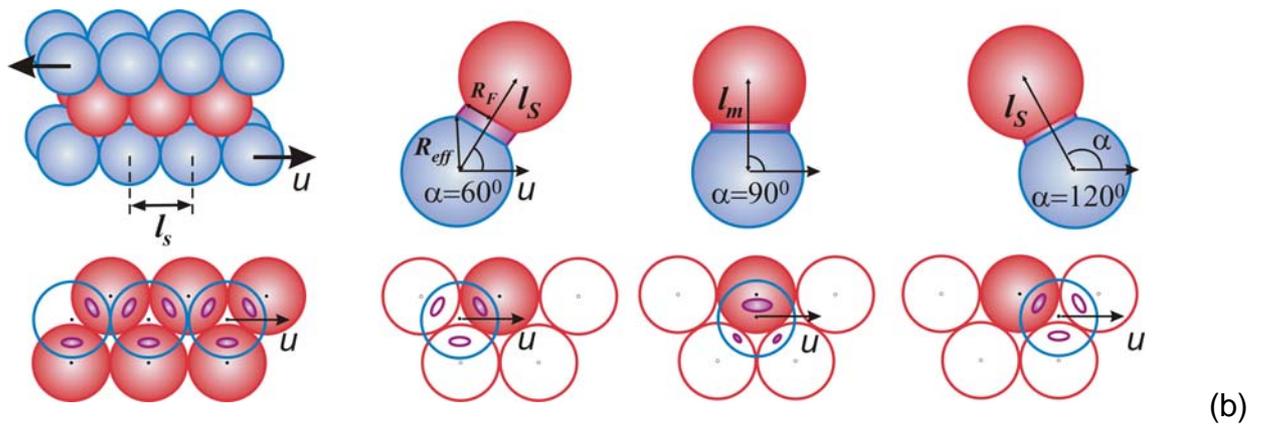

**FIG. 1.** (a) Illustration of the process of bubble-bubble friction with consecutive images of bubbles passing along each other in sheared foam. The images are taken by side-observation of foam, sheared between the parallel plates of rheometer. (b) Schematic presentation of the relative motion of the neighboring planes of bubbles in sheared foam (with relative velocity $u$), and of the process of film formation and disappearance between two bubbles, sliding along each other. The relative position of the bubbles is characterized by the distance $l(t)$ or by the angle, $\alpha(t)$, whereas $l_m$ denotes the distance of closest approach of the bubble centers at $\alpha = 90°$ (upper line: side view; bottom line: projection onto the plane of bubbles)

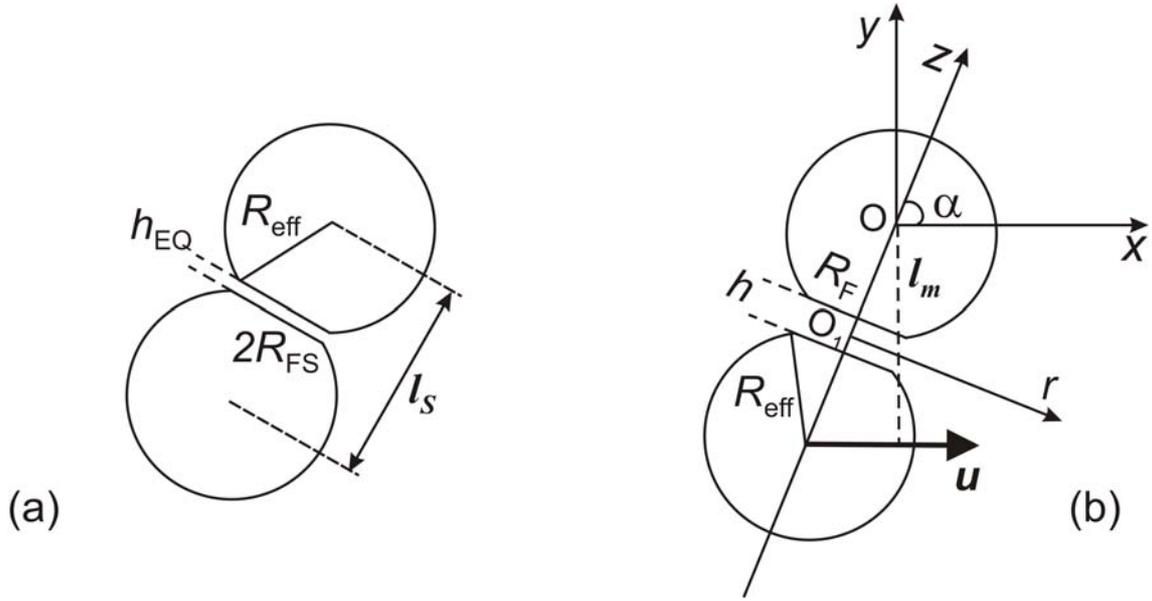

**FIG. 2.** Schematic presentation: (a) Two neighboring bubbles in static foam: $R_{eff}$ is the effective bubble radius, $l_S$ is the distance between bubble geometrical centers, $h_{EQ}$ is equilibrium film thickness, and $R_{FS}$ is film radius. (b) Sliding bubbles in sheared foam: $h$ is film thickness, $R_F$ is film radius, $l$ is center-to-center distance, $\alpha$ is running angle (see Fig. 1) and all these are functions of time. The coordinate system $Oxy$ used to describe the bubble relative position is fixed to the center of "immobile" upper bubble, whereas the center of the moving coordinate system $O_1 rz\varphi$, used in the description of film thinning, is located in the center of the planar film between the bubbles.

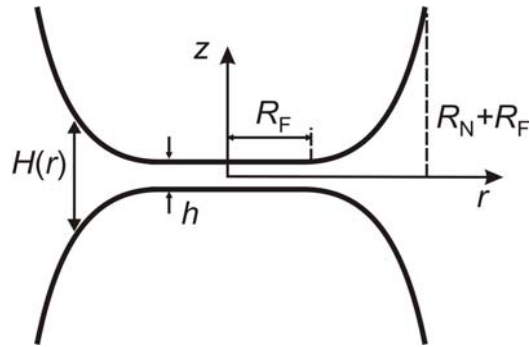

**FIG. 3.** Schematic presentation of transient foam film, $0 \leq r \leq R_F$, with the adjacent meniscus region $R_F \leq r \leq R_N + R_F$.

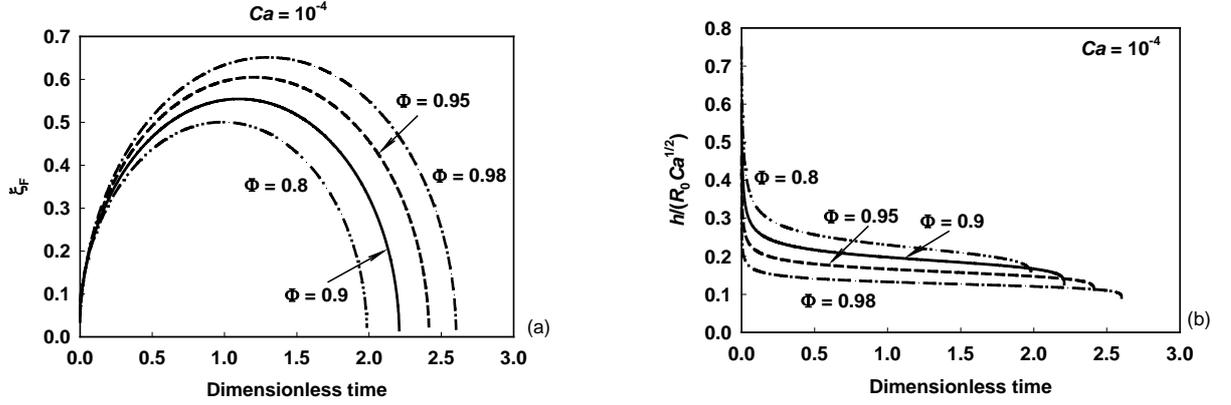

**FIG. 4.** Dimensionless (a) film radius, $\xi_F = R_F(\tilde{t})/R_0$, and (b) film thickness, $h(\tilde{t})/(R_0 Ca^{1/2})$, as functions of dimensionless time, $\tilde{t} = tu/R_0$, calculated for $Ca = 10^{-4}$ and different air volume fractions, $\Phi$. Equation (8) is used for $\tilde{P}_{OSM}$.

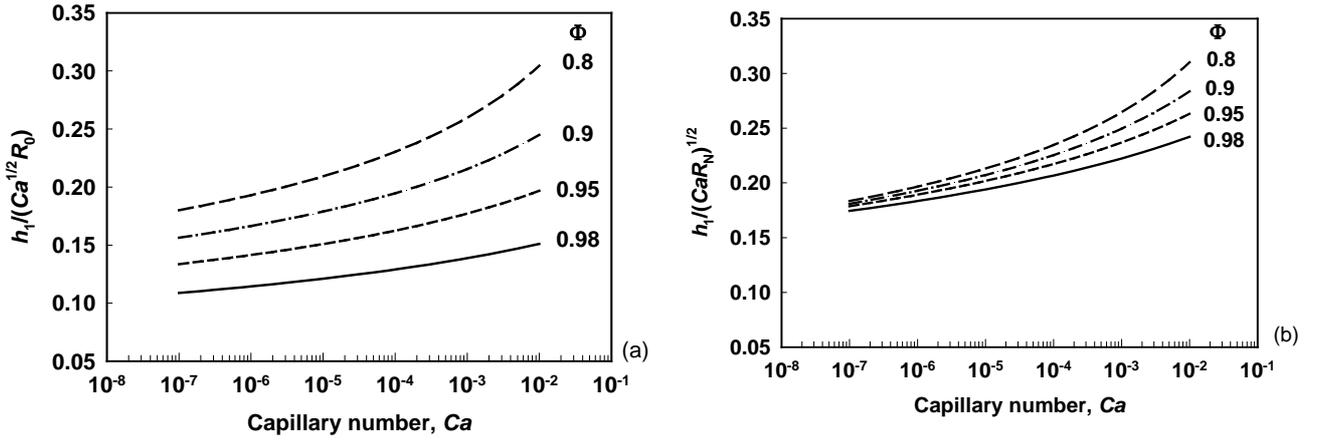

**FIG. 5.** (a) Dimensionless film thickness, $\eta_1/Ca^{1/2}$, as a function of the capillary number, $Ca$ for different air volume fractions; $\eta_1$ is defined as $\eta(l = l_m)$, which corresponds to $\tilde{t} = \tilde{x}_0 \approx 1$; (b) Scaled dimensionless film thickness, $h_1/(R_N Ca)^{1/2}$, as a function of $Ca$. Equation (8) is used for $\tilde{P}_{OSM}$.

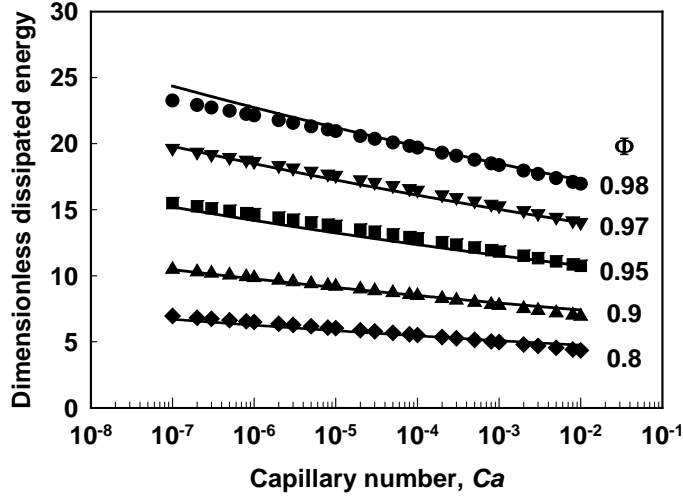

**FIG. 6.** Dimensionless energy dissipated in one film, $\tilde{E}_{DF} = E_{DF}/(R_0^2 \sigma \tilde{u}^{1/2})$, as a function of the capillary number, $Ca$, for different air volume fractions, $\Phi$. The symbols are numerical results, whereas the curves show the empirical interpolation, Eq. (48b). Equation (8) is used for $\tilde{P}_{OSM}$.

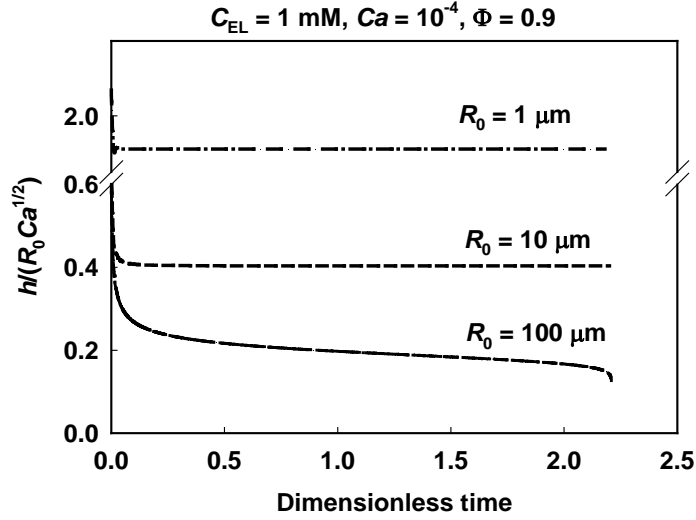

**FIG. 7.** Dimensionless film thickness as a function of dimensionless time, $\tilde{t} = tu/R_0$, for emulsions with different drop radii, $R_0$. The other parameters used for the calculations are: $A_H = 4 \times 10^{-21}$ J; $\Psi_S = 100$ mV; $\sigma = 5$ mN/m; T = 298 K; $C_{EL} = 1$ mM; $Ca = 10^{-4}$ and $\Phi = 0.9$. Equation (8) is used for $\tilde{P}_{OSM}$.

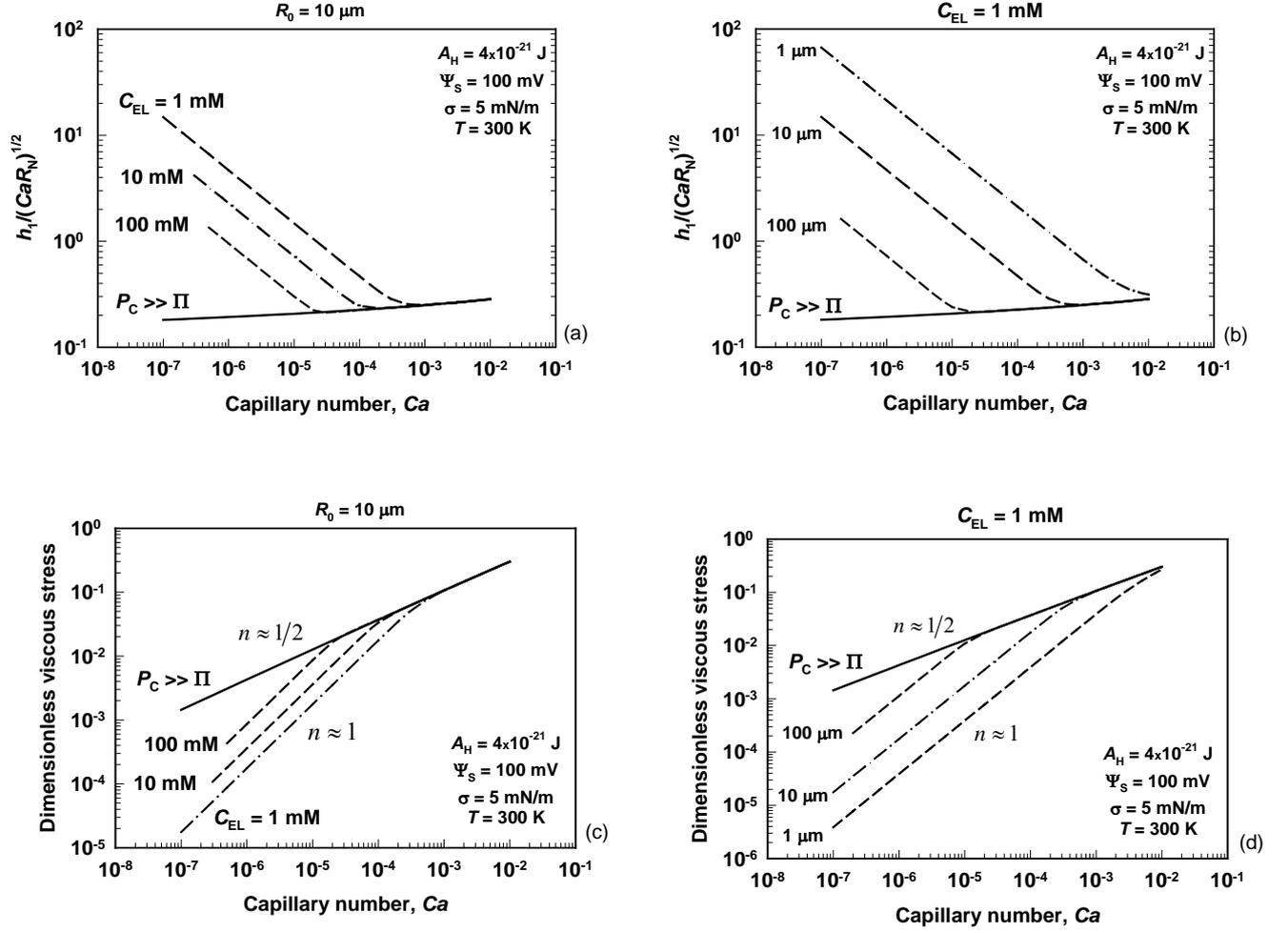

**FIG. 8.** (a,b) Dimensionless film thickness, $h_1/(R_N Ca)^{1/2}$, and (c,d) dimensionless viscous stress, $\tilde{\tau}_{VF} = \tau/(\sigma/R_0)$, as functions of the capillary number, $Ca$. (a,c) different electrolyte concentrations, (b,d) different drop sizes. The parameters used in the calculations are typical for emulsions stabilized by ionic surfactants: $A_H = 4\times 10^{-21}$ J, $\Psi_S = 100$ mV, $\sigma = 5$ mN/m, $T = 298$ K, and $\Phi = 0.9$. Equation (8) is used for $\tilde{P}_{OSM}$.

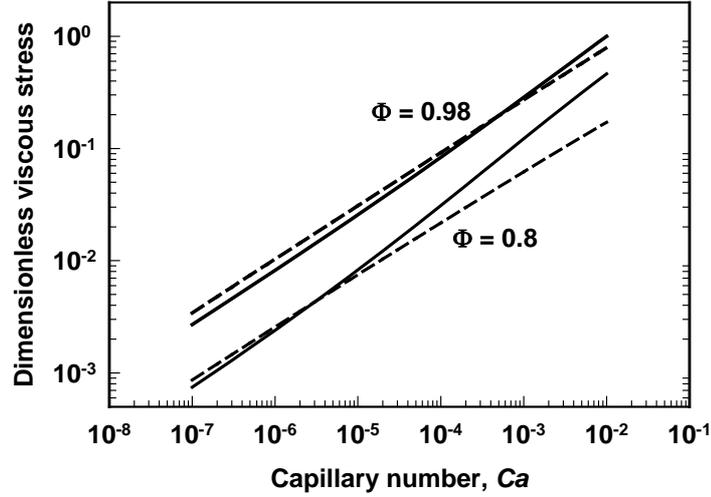

**FIG. 9.** Dimensionless viscous stress vs. capillary number. Solid curves – with account for the viscous friction in the meniscus region; dashed curves – with account for the friction in the films only. Equation (8) is used for $\tilde{P}_{OSM}$.

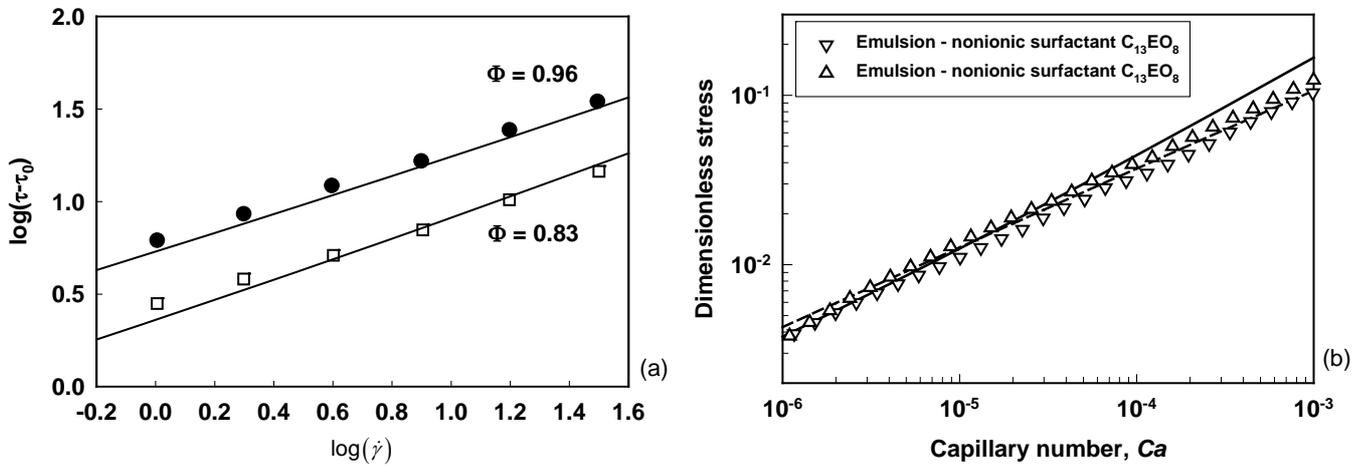

**FIG. 10.** Comparison of model predictions with experimental results (symbols) for oil-in-water emulsions: (a) Data from Ref. [4] for $\Phi = 0.83$ and $0.96$; (b) Our results obtained by parallel-plates rheometry, by using the method described in Ref. [26]. The solid curves in (a) and (b) are drawn according to Eq. (54), whereas the dashed curve in (b) is drawn with account for the dissipation in the film region only (all curves - without adjustable parameter).

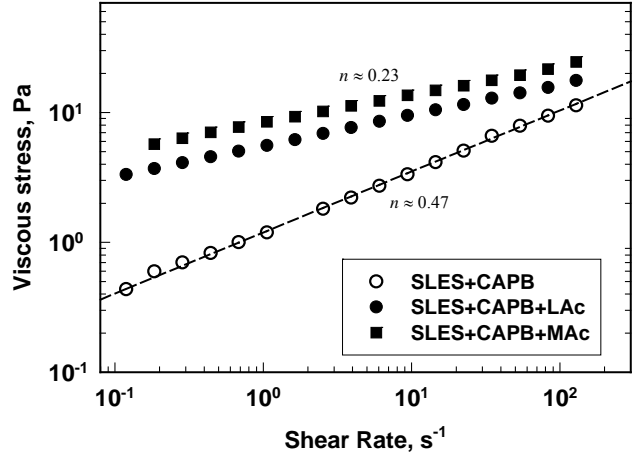

**FIG. 11.** Comparison of the theoretical predictions with experimental data for foams formed from surfactant solutions with different surface loss modulus, $G_{LS}$ (see section 6). The symbols are experimental data, the solid curve is drawn according to Eq. (38) (friction only in the foam films, no adjustable parameter).

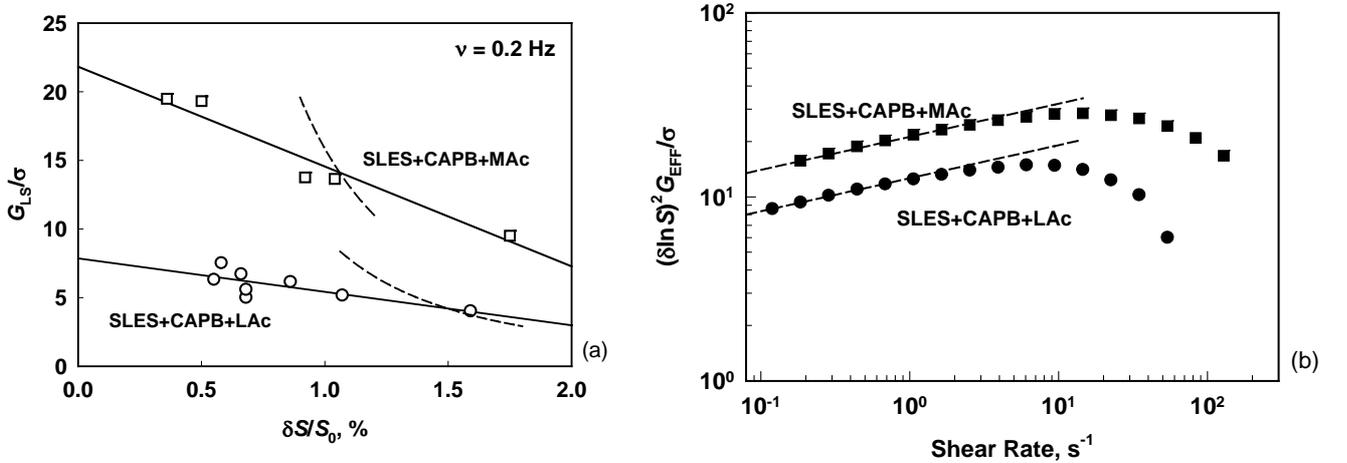

**FIG. 12.** (a) Comparison of the surface loss modulus, $G_{LS}$, measured by oscillating drop method (symbols and the solid lines, which are drawn to guide the eye) with the effective modulus $G_{EFF}$, determined from the foam viscous stress (dashed curves). From the cross-points one can determine the effective bubble deformation in the sheared foam, $\delta S/S_0$, which provides agreement between the calculated and the measured foam viscous stress, generated by the surface dissipation of energy, $\tau_{VS}$. (b) Effective surface loss modulus, $G_{EFF}$, determined from the foam viscous stress, after subtracting the contribution from the friction in the foam films and meniscus region, see Eq. (59). The lines are linear interpolations in the region of low shear rates, $\dot{\gamma} \leq 1 s^{-1}$.